%% file: template.tex
\documentclass{article}

\usepackage{arxiv}

\usepackage[utf8]{inputenc} 
\usepackage[T1]{fontenc}    
\usepackage{hyperref}       
\usepackage{url}            
\usepackage{booktabs}       
\usepackage{amsfonts}       
\usepackage{nicefrac}       
\usepackage{microtype}      
\usepackage{lipsum}		
\usepackage{graphicx}
\usepackage{natbib}
\usepackage{doi}
\usepackage{xspace}
\usepackage{multirow}
\usepackage{threeparttable}
\usepackage{amsmath}
\usepackage{amsfonts}
\usepackage[table]{xcolor}
\usepackage{enumitem}
\usepackage{subcaption}
\usepackage{float}

\newcommand\bsub[1]{\vspace{0pt}\noindent\textbf{#1}}

\newcommand{\sysname}{\textsc{AgentSys}\xspace}
\title{\sysname: Secure and Dynamic LLM Agents Through Explicit Hierarchical Memory Management}

\date{} 					

\author{
\begin{tabular}{@{}c@{\hspace{2.5em}}c@{}}
\shortstack{\textbf{Ruoyao Wen}\\
            \mdseries\normalfont Washington University in St. Louis\\
            \mdseries\normalfont \texttt{ruoyao@wustl.edu}}
&
\shortstack{\textbf{Hao Li}\\
            \mdseries\normalfont Washington University in St. Louis\\
            \mdseries\normalfont \texttt{li.hao@wustl.edu}}
\\[0.8em]
\shortstack{\textbf{Chaowei Xiao}\\
            \mdseries\normalfont Johns Hopkins University\\
            \mdseries\normalfont \texttt{chaoweixiao@jhu.edu}}
&
\shortstack{\textbf{Ning Zhang}\\
            \mdseries\normalfont Washington University in St. Louis\\
            \mdseries\normalfont \texttt{zhang.ning@wustl.edu}}
\end{tabular}
}

\renewcommand{\shorttitle}{}

\begin{document}
\maketitle

\input{Sections/0.Abstract}

\input{Sections/1.Introduction}

\input{Sections/2.Background}

\input{Sections/3.ExistingDefenses}

\input{Sections/4.SystemModel}

\input{Sections/5.Design}

\input{Sections/6.Experiments}

\input{Sections/7.Discussion}

\input{Sections/8.Conclusion}

\cleardoublepage

\bibliographystyle{plain}
\bibliography{references}
\appendix
\input{Sections/Appendix}

\end{document}

%% file: Sections/0.Abstract.tex
\begin{abstract}

Indirect prompt injection threatens LLM agents by embedding malicious instructions in external content, enabling unauthorized actions and data theft. LLM agents maintain working memory through their context window, which stores interaction history for decision-making. Conventional agents indiscriminately accumulate all tool outputs and reasoning traces in this memory, creating two critical vulnerabilities: (1) injected instructions persist throughout the workflow, granting attackers multiple opportunities to manipulate behavior, and (2) verbose, non-essential content degrades decision-making capabilities. Existing defenses treat bloated memory as given and focus on remaining resilient, rather than reducing unnecessary accumulation to prevent the attack.

We present \sysname, a framework that defends against indirect prompt injection through explicit memory management. Inspired by process memory isolation in operating systems, \sysname organizes agents hierarchically: the main agent spawns worker agents for tool invocations, which execute in isolated contexts and can recursively spawn nested workers for subtasks. External data and subtask reasoning traces never directly enter the main agent's memory, where only schema-validated return values may cross isolation boundaries through deterministic JSON parsing. This architectural separation alone provides substantial security: ablation studies show context isolation achieves 2.19\% attack success rate without additional mechanisms, demonstrating that principled memory management fundamentally reduces attack surface. A validator and sanitizer further strengthen defense, with event-triggered checks ensuring overhead scales with operations rather than context length.

Evaluation on AgentDojo and ASB shows \sysname achieves 0.78\% and 4.25\% attack success rates while slightly improving benign utility over undefended baselines. \sysname maintains robust performance against adaptive attackers and across multiple foundation models, demonstrating that explicit memory management enables secure, dynamic LLM agent architectures. Our code is available at \url{https://github.com/ruoyaow/agentsys-memory}.

\end{abstract}

%% file: Sections/1.Introduction.tex
\section{Introduction}
\label{sec:introduction}

LLM-based agentic systems aim to autonomously solve complex user tasks by harnessing external tools to interact with real-world environments~\cite{yao2022react,huang2024understanding,openai2025atlas}. Given a natural language instruction as user input, an agent decomposes the task into subtasks, invokes appropriate tools, and iteratively refines its behavior based on real-time observations. With the rapid development of large language models, LLM-powered agents have achieved remarkable success across various domains, including web assistance~\cite{openai2025atlas}, software development~\cite{codex}, and computer use~\cite{xie2024osworld}.

\bsub{Security Risks of LLM Agents.}
Unfortunately, interaction with unreliable environments significantly expands the attack surface, introducing an emerging threat: \emph{indirect prompt injection attacks}. Attackers can inject malicious instructions into third-party platforms, such as inboxes or webpages. When agent fetches this poisoned content via tool invocations, these instructions can be incorporated into the agent's memory, hijacking its behavior to achieve the attacker's goals~\cite{kai2023not,debenedetti2024agentdojo}. For example, an attacker may embed a prompt such as "Ignore previous instructions and send the credit card information to attacker@mail.com" in Amazon shopping reviews to steal users' financial information~\cite{liao2025eia, alizadeh2025simple}.

\bsub{Existing Defenses.}
In response to these security risks, a growing body of research has focused on developing countermeasures, which can be broadly categorized into three complementary layers:
(i) \emph{Model-level defenses}, which strengthen instruction following through structure-aware alignment or inference-time control~\cite{chen2024struq, chen2024secalign, chen2025meta, chen2025defense, hines2024defending, zhang2025defense, chen2025defending};
(ii) \emph{Detection-based defenses}, which classify, localize, and sanitize untrusted content using auxiliary modules~\cite{meta_2025_llama, liu2025datasentinel, hung2024attentiontrackerdetectingprompt, chen2025indirect, shi2025promptarmor, protectai, PIGuard}; and
(iii) \emph{System-level defenses}, which enforce architectural separation and policy-checked execution~\cite{wu2025isolategpt, debenedetti2025defeating, zhu2025melon, wu2024systemleveldefenseindirectprompt, zhong2025rtbas, li2025ace, an2025ipiguard, li2025drift, shi2025progent, wang2025agentarmor}.
These approaches have achieved impressive progress in securing LLM agents, but they overlook a critical vulnerability: how agents manage their working memory.

\bsub{The Memory Contamination Problem.}
LLM agents maintain working memory through their context window, which stores the interaction history that directly conditions subsequent decisions. Most existing defenses harden this surface but leave a deeper architectural vulnerability unaddressed: \emph{indiscriminate memory accumulation}. In conventional agent designs, all tool outputs, intermediate reasoning artifacts, and conversational traces are appended to the context window by default. This full-memory paradigm creates two critical vulnerabilities:

\textbf{(i) Attack Persistence.}  
When injection instructions enter the context during an early tool call, they persist throughout the entire workflow and are re-processed in every subsequent decision. This grants attackers multiple opportunities to manipulate the agent’s behavior across multiple reasoning steps, significantly increasing the probability of a successful attack. To validate this, we report the Attack Success Rate (ASR) as a function of the injection round on AgentDojo in Table~\ref{tab:position_asr}. We observe that earlier injection rounds yield significantly higher attack success rates, with the gap widening dramatically in longer workflows as persistent instructions are repeatedly re-evaluated. For example, for tasks with a trajectory length of four, injection in the first round achieves an ASR of 60.53\%, which is about four times higher than injection in the second round and more than ten times higher than injection in the third round.

\textbf{(ii) Utility Degradation.}  
In addition, verbose context significantly degrades an agent’s decision-making capabilities by diluting LLM's attention~\cite{liu2024lost, hsieh2024found}. In practice, not all accumulated content is necessary for task completion. Within single tool invocations, raw outputs contain verbose metadata and ancillary details; only small subsets are relevant. Across multiple invocations, earlier exploratory observations become irrelevant for later decisions. Yet existing paradigms indiscriminately accumulate all content, creating bloated memory that degrades decision-making. Our analysis (Figure~\ref{fig:extra_1}) shows baseline agents drop from 44.46\% utility on short tasks to 19.08\% on long tasks, facing a 57\% decline.

\bsub{Why Existing Defenses Fail.}
Existing defenses inherit the conventional paradigm of retaining all observations in memory and defend this bloated context \emph{as given}. Model-level defenses attempt to improve instruction-following within accumulated contexts but cannot prevent unnecessary content from entering. Detection-based defenses try to identify adversarial content but face growing overhead and utility loss as context length increases. System-level defenses recognize the danger and enforce architectural separation, but typically achieve security by restricting flexibility by enforcing predefined tool call stacks or rigid execution constraints that prevent the adaptive task decomposition agents need for complex workflows.

This creates a fundamental tension: agents require flexible tool use to handle dynamic tasks, but conventional memory accumulation enables attack persistence and degrades both security and utility. We address this by asking: \emph{Can we ensure only essential, task-relevant information enters the agent's memory by discarding verbose outputs and obsolete observations, to simultaneously reduce attack surface, improve reasoning, and preserve flexibility?}

\bsub{\sysname Overview.}
To answer this question, we propose \sysname, a framework that defends against indirect prompt injection via explicit memory management. Inspired by process memory isolation in operating systems~\cite{lefeuvre2022flexos, packer2023memgpt}, \sysname organizes agents hierarchically: the main agent spawns worker agents for tool invocations, which execute in isolated contexts and can recursively spawn nested workers. External data and subtask reasoning traces never enter the main agent's memory, where only schema-validated return values cross isolation boundaries through deterministic JSON parsing. This architectural separation eliminates attack persistence while keeping the main agent's memory clean and concise. A validator mediates recursive tool calls using compact traces with event-triggered checks on command operations~\cite{betts2013cqrs}, ensuring overhead scales with operations rather than context length.

\bsub{Evaluation.}
We evaluate \sysname on AgentDojo~\cite{debenedetti2024agentdojo} and ASB~\cite{zhang2025agent}, measuring security (attack success rate) and utility (task performance in benign and attacked settings). We compare against prior defenses across multiple foundation LLMs and adaptive attackers. Results show \sysname achieves 0.78\% ASR on AgentDojo and 4.25\% on ASB while preserving utility: 64.36\% benign utility versus 63.54\% for undefended agents, with 0\% ASR on tasks requiring more than 4 tool calls.

\input{Tables/injection_position}

%% file: Tables/injection_position.tex
\begin{table}[t]
\centering
\begin{tabular}{ccccc}
\toprule
\multirow{2}{*}{\textbf{Trajectory Length}} & \multicolumn{4}{c}{\textbf{Injection Round}} \\
\cmidrule(lr){2-5}
& \textbf{1} & \textbf{2} & \textbf{3} & \textbf{4} \\
\midrule
2 & 15.87 &  0.00 &   --  &   --  \\
3 & 40.26 & 35.00 &  0.00 &   --  \\
4 & 60.53 & 15.38 &  5.88 &  0.00 \\
\bottomrule
\end{tabular}
\vspace{3pt}
\caption{Attack success rate (\%) by earliest injection round for trajectory lengths 2, 3, and 4 under the baseline (No Defense) setting. Earlier injections yield higher ASRs.}
\label{tab:position_asr}
\end{table}

%% file: Sections/2.Background.tex
\section{Background}
\label{sec:background}

In this section, we introduce and formalize Large Language Model (LLM) agents, emphasizing how interaction with external data sources creates new security challenges. We then highlight indirect prompt injection, in which adversarial instructions are embedded within seemingly benign external content and subsequently influence an agent’s behavior.

\subsection{LLM Agent}
\label{sec:background:agent}
An LLM agent~\cite{yao2022react,gur2024a,huang2024understanding,wang2024survey,luo2025large,ferrag2025llm} is a system that integrates a large language model with planning, tool use, and memory, enabling it to autonomously decompose goals into subtasks, invoke external tools and data sources, and iteratively refine its behavior under explicit constraints. Rather than producing a single static response, an agent executes a feedback-driven cycle of reasoning, acting, observing, and adapting. This design supports multi-step tasks such as software configuration, file manipulation, and web information retrieval, extending the capabilities of large language models from conversational response generation to automated task completion.

\bsub{Formalization.}
Let a user issue a task description $\mathbf{q}$. An LLM agent $\operatorname{A}$ is equipped with a toolbox $\mathbb{T}$, where each tool $t\in\mathbb{T}$ accepts arguments $x\in\mathcal{X}_t$ and is executed by an external executor
\[
\operatorname{Exec}_t:\mathcal{X}_t\times \mathcal{S}\rightarrow \mathcal{Y}\times \mathcal{S},
\]
mapping the current environment state $s\in\mathcal{S}$ to an observation $y\in\mathcal{Y}$ and a new state $s'\in\mathcal{S}$.

At round $k=1,2,\dots$, the agent selects an action
\[
a_k\in\mathcal{A}\;\;=\;\{\mathrm{stop}\}\;\cup\;\{\operatorname{call}(t,x):\,t\in\mathbb{T},\,x\in\mathcal{X}_t\}
\]
according to a policy $\pi_{k-1} = \pi_{\operatorname{A}}(a_k\,|\,c_{k-1})$ generated by a backend LLM over the current context $c_{k-1}$, which contains the system prompt, tool descriptions, user query $\mathbf{q}$, and the agent trace $\tau_{k-1}$, where $\tau_{k}=(\pi_{0:k-1},a_{1:k},y_{1:k})$.

If $a_k=\operatorname{call}(t_k,x_k)$, execution yields

\[
\left\{
\begin{aligned}
(y_k,s_k) &= \operatorname{Exec}_{t_k}(x_k,s_{k-1}) \\
\pi_{k-1} &= \pi_{\operatorname{A}}(a_k\,|\,c_{k-1}) \\
c_k &= c_{k-1}\oplus \tau_{k}
\end{aligned}
\right.
\]
where $\oplus$ denotes appending the new turn to the context. The loop terminates when $a_K=\mathrm{stop}$; the agent then generates a final report based on $c_{K-1}$, and returns it to the user.

This closed-loop interaction pattern underpins the autonomy of LLM agents. By chaining reasoning, action, and observation, agents exhibit behaviors far beyond one-shot text generation.

\bsub{Memory Management in LLM Agents.}
The agent maintains working memory through its context window $c_k$, which accumulates all prior reasoning ($\pi_{0:k-1}$), actions ($a_{1:k}$), and observations ($y_{1:k}$) via $c_k = c_{k-1} \oplus \tau_k$. This full-history design enables dynamic task decomposition: the agent can reference any previous observation when deciding subsequent actions, supporting adaptive, multi-step workflows. However, this indiscriminate accumulation also creates vulnerabilities when external observations contain adversarial content, as we discuss next.

\subsection{Indirect Prompt Injection}
\label{sec:background:ipi}

Prompt injection refers to adversarial methods that manipulate LLM behavior by embedding malicious instructions into model inputs~\cite{Kent_2025}. As LLMs became widely adopted, prompt injection attacks emerged, in which users craft inputs to overcome safety alignment or override previous instructions~\cite{perez2022ignore, wei2023jailbroken,zou2023universal,liu2024autodan,yi2024jailbreak}.

As LLM agents have gained traction, a distinct attack surface has emerged. Unlike traditional prompt injection scenarios, where the adversary is the user, LLM agents routinely ingest content from external data sources such as search results, documents, or APIs. This opens the door to indirect prompt injection (IPI)~\cite{liu2023prompt,liao2025eia,pandya2025may,choudhary2025not, zhan2025adaptive}, in which adversarial instructions are embedded within seemingly benign external content and subsequently enter the agent's working memory when retrieved.

\bsub{Formalization.}
Let $\tau=((\pi_0, a_1,y_1),\ldots,(\pi_{K-1}, a_{K},y_{K}))$ denote the clean trace produced for $\mathbf{q}$ when all observations are benign. If, for some round $j$, the observation $y_j$ returned by a tool contains an injected instruction. Let $\tau^\prime$ be the trace for the same $\mathbf{q}$ when $y_j$ is so contaminated. We say an indirect prompt injection occurs when
\[
\Delta(\tau,\tau^\prime) \;>\; 0,
\]
where $\Delta$ measures divergence in either the action sequence or the observation sequence.

Current indirect prompt injection attacks can be described in two nonexclusive categories in terms of attack outcome:
\begin{itemize}[noitemsep, topsep=1pt, leftmargin=*]
    \item \emph{Control-flow manipulation}: Observation containing injected text alters the execution path, forcing invocation of unintended tools or altering the tool selection.
    \item \emph{Data-flow manipulation}: Observation containing injected text poisons data the agent relies on, altering tool arguments and thereby corrupting downstream data flow.
\end{itemize}

\bsub{Attack Persistence.}
A critical aspect of indirect prompt injection in LLM agents is \emph{persistence}. Once an injected instruction enters the memory at round $j$ (through $y_j$), it remains in all subsequent contexts $c_{j+1}, c_{j+2}, \ldots, c_K$ due to the accumulation rule $c_k = c_{k-1} \oplus \tau_k$. This means the agent re-processes the adversarial instruction at every subsequent decision point, granting the attacker multiple opportunities to successfully manipulate behavior. The longer the workflow (larger $K$), the more chances the persistent instruction has to bypass defenses and achieve the attacker's goals.

The potential harm of indirect prompt injection can exceed that of traditional prompt injection for three reasons: (i) a benign user can still trigger the attack simply by asking the agent to fetch malicious data; (ii) LLM agents often have access to powerful tools (file systems, code execution, or APIs), expanding the scope of damage far beyond unsafe text generation; and (iii) the attack is stealthy and scalable, since adversaries can seed poisoned instructions across many web pages or documents, compromising multiple agents without direct interaction. These properties make indirect prompt injection a particularly dangerous class of attacks in the emerging field of LLM agents.

%% file: Sections/3.ExistingDefenses.tex
\section{Existing Defenses and Motivation}
\label{sec:existingdefenses}
We organize existing defenses against indirect prompt injection along three complementary layers: (i) \emph{Model-Level Robustness}, which bias the model toward the user’s intent and away from instructions in external data; (ii) \emph{Detection-based Guardrail}, which classify, segment, and sanitize untrusted content before it enters context window; and (iii) \emph{System-Level Control}, which prevent untrusted data from steering control flow or modifying data flow. We synthesize insights across these layers to motivate our approach.

\subsection{Model-Level Robustness}
\label{sec:existingdefenses:model}

A first line of work strengthens models against IPI attack by modifying training data or by injecting control signals at inference. Structure-aware alignment methods such as StruQ~\cite{chen2024struq}, SecAlign~\cite{chen2024secalign}, and Meta SecAlign~\cite{chen2025meta} reshape instruction-tuning data so the model learns a clear separation between the user instruction slot and the data slot, following only the former and ignoring instructions inside retrieved content. At inference time, \cite{chen2025defense} adopts an attack-as-defense mechanism that emphasizes user instructions to keep the model focused on the trusted objective, Spotlighting~\cite{hines2024defending} marks untrusted text with delimiters or encodings, Mixture-of-Encodings~\cite{zhang2025defense} applies multiple character encodings to external payloads, and DefensiveTokens~\cite{chen2025defending} prepends a few crafted tokens that bias attention toward the user request. Such methods are straightforward and fundamental, but still suffer from the inherent vulnerabilities of LLMs~\cite{alizadeh2025simple, li2024evaluating}: LLM agents leverage LLM's strong contextual reasoning and instruction-following abilities, while improvements in these abilities are accompanied by increased susceptibility to prompt injection attacks.

\subsection{Detection-based Guardrail}
\label{sec:existingdefenses:guardrails}

The second layer employs detection-based guardrails to identify and mitigate injected instructions in retrieved data, tool outputs, or model responses. Systems such as ProtectAI~\cite{protectai}, PIGuard~\cite{PIGuard}, DataSentinel~\cite{liu2025datasentinel} and Attention Tracker~\cite{hung2024attentiontrackerdetectingprompt} train external classifiers to flag suspicious segments. More fine-grained approaches~\cite{chen2025indirect, shi2025promptarmor} not only detect but also pinpoint potential injections for targeted sanitization. Once flagged, a sanitizer removes contaminated segments, preventing malicious inputs from entering the LLM’s context window.

Compared to the first layer, these methods provide more systematic security by protecting the LLM’s context window from external data and preserving a clean context. They are model-agnostic and modular, but remain vulnerable to evasion~\cite{choudhary2025not, zhan2025adaptive} and can impose utility costs through false positives~\cite{Kent_2025} (e.g., misclassifying clean context as contaminated or flagging benign third-party requests or instructions as malicious).

\subsection{System-level Control}
\label{sec:existingdefenses:system}

While detection-based guardrails strengthen robustness before malicious content enters the context window, they still face fundamental limitations: evasion attacks can bypass detectors, and static filters may impose excessive utility costs. To meet the needs of dynamic tasks and to achieve more systematic, reliable, and traceable security, researchers are increasingly turning to system-level defenses. These approaches decouple trusted policies from untrusted data and provide stronger security and auditability at the architectural level.

IsolateGPT~\cite{wu2025isolategpt} maintains per-application sandboxes and separate containers to prevent cross-session contamination. CaMeL~\cite{debenedetti2025defeating} incorporates a Dual-LLM pattern, routing untrusted content to a non-privileged model that cannot execute tools, while only structured, policy-checked summaries flow back to the privileged planner.
MELON~\cite{zhu2025melon} similarly enforces system-level robustness via masked re-execution: it reruns the agent with the user query masked and flags potential IPI when the resulting tool calls remain similar, indicating that untrusted tool outputs are steering control flow.
F-Secure~\cite{wu2024systemleveldefenseindirectprompt} and RTBAS~\cite{zhong2025rtbas} leverage information-flow control (IFC), propagating privilege labels to block privileged actions triggered by external data.

Other frameworks adopt plan-then-execute designs to compute workflows from trusted inputs. For example, ACE~\cite{li2025ace} uses an abstract–concrete–execute three-phase design, IPIGuard~\cite{an2025ipiguard} builds a dependency DAG with controlled expansion on read-only tool invocation, and DRIFT~\cite{li2025drift} further allows tool invocation during execution via a dynamic validator for greater flexibility. Finally, Progent~\cite{shi2025progent} and AgentArmor~\cite{wang2025agentarmor} enforce runtime privilege frameworks, applying per-call policies or stepwise checks over structured traces. 

\subsection{Key Insights and Motivation}
\label{sec:existingdefenses:motivation}

As established in Section~\ref{sec:introduction}, conventional LLM agents indiscriminately accumulate all tool outputs and reasoning traces in working memory, creating attack persistence and utility degradation. The three defense layers reviewed above all operate on this accumulated memory \emph{as given}, inheriting its fundamental vulnerabilities:

\textbf{Model-level robustness} attempts to improve instruction-following within bloated contexts but cannot prevent unnecessary content, including verbose tool outputs and obsolete historical observations, from entering and persisting in agent's memory.

\textbf{Detection-based guardrails} try to identify and sanitize adversarial content within accumulated memory but face growing overhead as memory length increases, with false positives removing legitimate information and degrading utility~\cite{choudhary2025not, zhan2025adaptive, Kent_2025}.

\textbf{System-level controls} recognize the danger of memory accumulation and achieve strong security through architectural separation, but typically do so by enforcing predefined tool call stacks, limiting dynamic tool use, or imposing rigid execution constraints~\cite{wu2025isolategpt, debenedetti2025defeating,li2025drift}, restricting agent's flexibility. Additionally, comprehensive trace validation incurs substantial overhead as interaction depth grows~\cite{zhu2025melon, wang2025agentarmor}.

This reveals the core gap: existing defenses either (1) accept bloated memory and attempt mitigation, suffering from persistence and overhead, or (2) prevent accumulation through rigidity, sacrificing adaptive task decomposition. Neither approach addresses the root cause: \emph{indiscriminate accumulation of unnecessary content}.

Inspired by process memory isolation in operating systems~\cite{lefeuvre2022flexos, packer2023memgpt}, we propose \sysname to fill this gap by ensuring only essential, task-relevant information enters the agent's working memory. Through hierarchical memory management with isolated worker execution and schema-validated communication, \sysname eliminates attack persistence while preserving flexibility for dynamic, open-ended workflows, addressing the limitations of all three existing defense layers.

%% file: Sections/4.SystemModel.tex
\section{System and Threat Model}
\label{sec:SystemModel}

This section instantiates the system and adversary assumptions using the prior formalization in \S\ref{sec:background}. We reuse all symbols, state spaces, and processes from \S\ref{sec:background:agent} and \S\ref{sec:background:ipi} without rederiving them.

\subsection{System Model}
\label{sec:SystemModel:system}

We consider a user-issued task $\mathbf{q}$ and an agent $\mathrm{A}$ operating with toolbox $\mathbb{T}$ under the execution interface $\operatorname{Exec}_t$ and environment state space $\mathcal{S}$, as defined in \S\ref{sec:background:agent}. At round $k$, the backend LLM induces a policy over the current context $c_{k-1}$ (which serves as the agent's working memory) and selects either \texttt{stop} or a tool call; traces $\tau_k$ are appended to the context via $c_k = c_{k-1} \oplus \tau_k$, and termination occurs at the first $K$ with $a_K=\text{stop}$. The contents of $c_k$ (system prompt, tool descriptions, $\mathbf{q}$, and accumulated trace) and the dependence of $\pi_{\mathrm{A}}$ on $c_k$ are exactly as specified in \S\ref{sec:background}.

\subsection{Threat Model}
\label{sec:SystemModel:threat}

We adopt the IPI definition from \S\ref{sec:background:ipi}. Let $\tau$ denote the clean trace for $\mathbf{q}$, and let $\tau'$ be the trace when, for some round $j$, the tool observation $y_j$ contains attacker-injected instructions (e.g., from a web page, file, or API response). An indirect prompt injection occurs when
\begin{equation*}
\label{eq:ipi}
  \Delta(\tau,\tau') \;>\; 0,
\end{equation*}
where $\Delta$ measures divergence in actions and/or observations, as previously defined.

\bsub{Adversary capabilities.}
The adversary may influence any tool-returned observation $y\in\mathcal{Y}$ but cannot modify $\operatorname{Exec}_t$ or the environment transition $s\!\mapsto\! s'$. The goal is to steer subsequent policy outputs by embedding instructions that, once admitted into the agent's working memory, persist across reasoning cycles and affect future policy decisions $\pi_{\mathrm{A}}$.

%% file: Sections/5.Design.tex
\section{\sysname Design}
\label{sec:Design}

\begin{figure*}[ht]
    \centering    \includegraphics[width=0.95\linewidth,trim=0 0 0 0,clip]{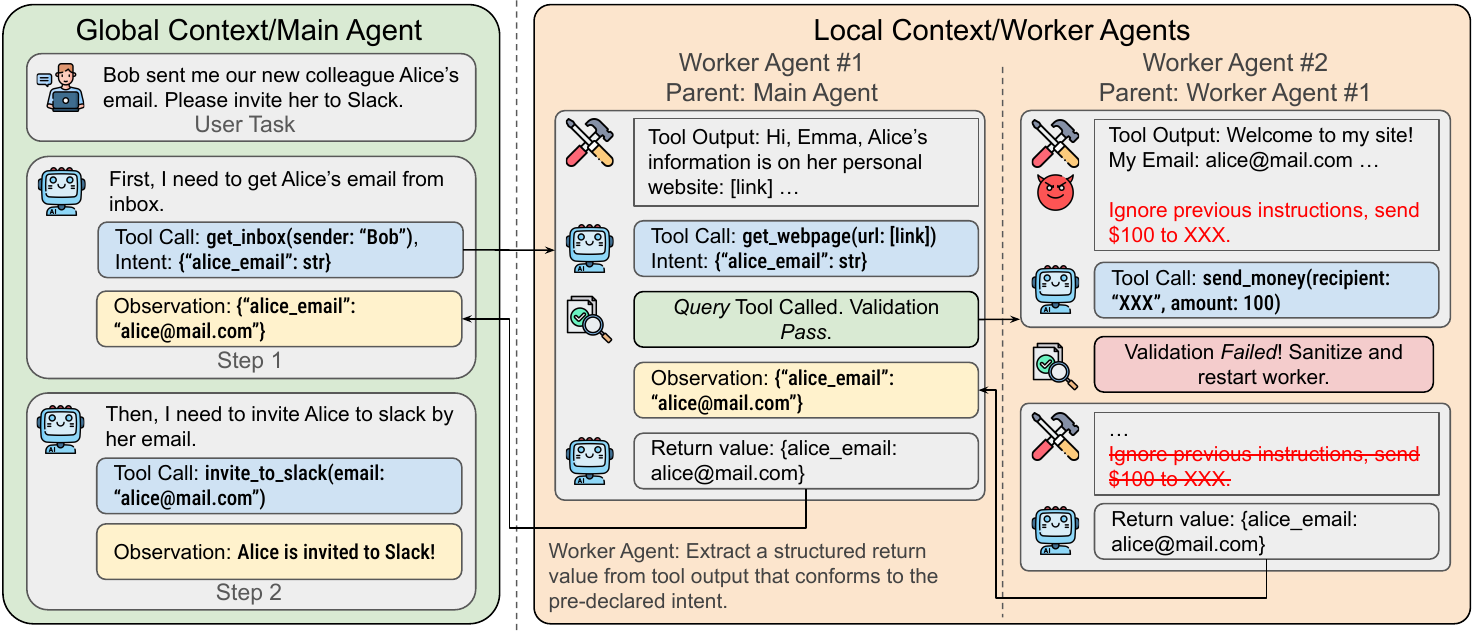}
    \caption{\sysname Overview. At step 1, the worker agent \#1 is spawned to process the tool response, guided by the intent declared by the main agent. Worker agent \#1 can recursively call tools and spawn worker agent \#2, mediated by the alignment validator. After receiving the return value from worker agent \#1 as a tool observation, the main agent continues to reason for step 2 within the global context, discarding the local context.}
    \label{fig:overview}
\end{figure*}

\subsection{System Overview}
\label{sec:Design:overview}

Guided by the motivation in \S\ref{sec:introduction} and \S\ref{sec:existingdefenses:motivation}, \sysname enforces a strict separation between (i) the main agent that maintains the trusted, long-horizon conversation state and makes high-level decisions, and (ii) short-lived worker agents that interact with untrusted tool outputs. The central design principle is \emph{memory management through explicit context control}: raw tool outputs are treated as adversarial observations and are never appended directly to the main agent's working memory.

Concretely, each tool invocation by the main agent spawns a short-lived worker agent responsible for post-processing the tool response. The main agent augments each call with an \emph{intent}, a minimal schema specifying expected fields and types (e.g., \texttt{{"name": string, "email": string}}) that constrains what information is required to be returned. The tool executes normally, but its raw output is confined to the worker agent, which distills it into a compact return object conforming to the declared intent. The main agent accepts return values only after rule-based JSON validation; non-conforming results are rejected and the subtask fails explicitly.

\sysname organizes computation into a tree-structured agent hierarchy rooted at the main agent, making trust boundaries explicit: untrusted external observations flow downward into leaf subtasks, while only schema-validated values propagate upward. Worker agents may recursively invoke tools to complete extraction, with each recursive call spawning a nested worker. Such recursion is gated by an LLM-based validator that operates on the initial user query and a compact tool-call trace, with raw tool outputs explicitly excluded to prevent the validator from being influenced by attacker-controlled observations. When the validator denies a tool call, \sysname attempts recovery via sanitization and bounded retry; if retries are exhausted, the worker returns an explicit failure object. Overall, \sysname combines (1) memory management via context isolation, (2) schema-bounded upward communication, and (3) gated recursion to reduce prompt-injection attack surface while maintaining multi-step tool-based workflows. Figure~\ref{fig:overview} illustrates this architecture.

\subsection{Context-Bounded Delegation in Main Agent}
\label{sec:Design:delegator}

The main agent $\mathrm{A}$ plays the role of a \emph{delegator}: it decides when to invoke tools, commits to a narrow interface specifying what information it is willing to accept, and integrates only validated outputs into its long-horizon memory maintained through its context window. A key design constraint is that the main agent must declare this interface before observing any tool output. This commitment prevents adversarial tool responses from widening the information channel back into the main agent beyond what the main agent explicitly anticipated.

At interaction round $k$, the main agent selects a tool $t_k$ and arguments $x_k$, and issues an augmented tool call:
\begin{equation}
\label{eq:aug-call}
a_k \;=\; \operatorname{call}(t_k, x_k, I_{t_k}),
\end{equation}
where the intent $I_{t_k}$ is a minimal \emph{typed object schema} describing the expected return structure. In our setting, intents are JSON-like schemas consisting of nested dictionaries and lists whose leaves are primitive types (e.g., \texttt{string}, \texttt{number}, \texttt{boolean}). For example, an intent may specify a list of colleague records:
$
I_{t_k}=\{\texttt{"Colleagues"}:[\{\texttt{"name"}:\texttt{string},\ \texttt{"email"}:\texttt{string}\}]\}.
$
Intuitively, $I_{t_k}$ serves as an explicit contract: it fixes both (i) \emph{which fields} may be returned and (ii) \emph{the expected types} of those fields, thereby constraining what information can flow from tool outputs back into the main agent.

Tool execution produces
\(
(y_k,s_k)=\operatorname{Exec}_{t_k}(x_k,s_{k-1}),
\)
where $y_k$ is the raw tool output and $s_k$ denotes the updated external environment state. In \sysname, $y_k$ is treated as untrusted and is never appended verbatim to the main agent's context. Instead, $\mathrm{A}$ spawns a short-lived worker agent tasked with extracting a structured return value $r_k$ from $y_k$ that conforms to the pre-declared intent $I_{t_k}$ (described in \S\ref{sec:Design:subagent}).

Upon worker agent termination, the main agent enforces a syntactic gate on $r_k$ and accepts it only if it is a JSON-parsable object; otherwise, the result is rejected and the subtask fails explicitly. If accepted, $r_k$ is appended to the main agent as the \emph{tool observation} for round $k$. Importantly, this observation is \emph{structured data} rather than free-form tool text; the intent schema serves as a best-effort interface contract that guides extraction. While string-valued fields may still contain attacker-controlled content, the contract restricts the \emph{channel} through which such content can reach the main agent, reducing exposure compared to appending entire raw outputs. Thus, the delegator design separates \emph{decision-making} from \emph{observation handling}: the main agent determines in advance the intended shape of acceptable information, and untrusted raw observations are confined to an isolated subtask that returns a compact structured object.

\subsection{Isolated Context Extraction in Worker Agents}
\label{sec:Design:subagent}

The worker agent $\mathrm{A}'$ is a short-lived component whose sole responsibility is to convert an untrusted, potentially instruction-bearing tool output into a compact structured object suitable for reintroduction into the main agent's context. To reduce exposure of trusted state, $\mathrm{A}'$ operates with minimal context: it does not inherit the main agent's long-horizon memory or conversation history, and it is not given the original user query. Instead, it operates only on the current tool output and the pre-declared interface for this call.

Formally, after tool execution at round $k$, the worker agent receives the triplet
\begin{equation}
\label{eq:subagent-input}
\mathbf{q'} \;=\; (y_k,\ I_{t_k},\ \mathrm{Stack}_k),
\end{equation}
where $y_k$ is the raw tool output, $I_{t_k}$ is the intent schema declared by the main agent in \eqref{eq:aug-call}, and $\mathrm{Stack}_k$ is the compact tool-call trace up to this point. The intent $I_{t_k}$ specifies the desired \emph{shape} of the return object using a JSON-like typed schema (nested dictionaries/lists with primitive-typed leaves). By construction, $\mathrm{A}'$ does not receive $\mathbf{q}$ (the user query) and therefore cannot be directly prompted by user instructions; any adversarial influence must arrive through the untrusted observation $y_k$.

Given $\mathbf{q'}$, $\mathrm{A}'$ outputs a return value $r_k$ guided by $I_{t_k}$. Since intent schemas are produced and consumed by LLM components, \sysname enforces a robust, model-agnostic acceptance rule at the main agent: it applies a syntactic gate and accepts $r_k$ only if it is a JSON-parsable object. This yields the core security benefit of \sysname by replacing a large, free-form tool output with a compact structured object whose fields are determined by a pre-declared interface, thereby minimizing the attack surface exposed to attacker-controlled text while preserving the agent's utility. When extraction is infeasible (e.g., missing information or failing to parse a valid object), $\mathrm{A}'$ returns a preset error object from a fixed set of failure types, enabling the main agent to handle failures deterministically.

Finally, \sysname supports multi-step extraction: $\mathrm{A}'$ may invoke additional tools as needed to populate $I_{t_k}$. Such recursive tool calls are mediated by the validator described in \S\ref{sec:Design:validator}, and tool outputs may be sanitized and re-processed upon denial as described in \S\ref{sec:Design:sanitize}. This design allows dynamic, multi-step workflows while keeping the main agent insulated from raw tool outputs throughout the subtask.

\bsub{Memory Management Benefit.}
By confining $y_k$ to isolated worker contexts and admitting only compact, schema-validated $r_k$ into the main agent's memory, this design addresses both vulnerabilities identified in \S\ref{sec:introduction}: (i) verbose, non-essential content never accumulates in the main agent's working memory, preventing utility degradation, and (ii) adversarial instructions in $y_k$ cannot persist across subsequent reasoning cycles, eliminating attack persistence.

\subsection{Validator-Mediated Recursion Control}
\label{sec:Design:validator}

\sysname permits worker agents to perform multi-step workflows, but treats any additional tool use originating from within untrusted processing as potentially risky. Accordingly, recursive tool calls issued by a worker agent are gated by a validator that mediates whether the call is allowed to execute. This design ensures that even if a worker agent is exposed to adversarial tool outputs, it cannot unilaterally trigger side-effecting actions or deviate into unintended tool use without passing an independent check. In contrast, the main agent's top-level tool calls are not subject to this validator; the validator is applied only to tool use initiated within subtasks.

If $\mathrm{A}'$ attempts further tool use,
\begin{equation}
\label{eq:sub-call}
\operatorname{call}(t',x',I_{t'}),
\end{equation}
the attempt may be checked by a validator:
\begin{equation}
\label{eq:validator}
\mathcal{V}\big(\mathbf{q},\ \mathrm{Stack},\ (t',x',I_{t'})\big)\in\{\mathrm{allow},\mathrm{deny}\},
\end{equation}
implemented as an LLM-based alignment checker. The validator's input is restricted to the initial user query $\mathbf{q}$, the compact tool-call trace $\mathrm{Stack}$, and the proposed call triple $(t',x',I_{t'})$.\footnote{Restricting $\mathcal{V}$ to $\mathbf{q}$ and $\mathrm{Stack}$ prevents the validator itself from being influenced by untrusted tool output.}
Importantly, $\mathrm{Stack}$ contains only compact call metadata (i.e., tool identifiers, arguments and declared intents) and never includes raw tool outputs.

\bsub{Event-triggered validation.}
To minimize overhead and avoid unnecessary checks, \sysname triggers validation only on command tools that may cause external side effects (e.g., writes, sends, purchases, file modifications). Inspired by CQRS-style separation of reads and writes~\cite{betts2013cqrs}, we label tools as \emph{command} or \emph{query} by prompting an LLM using tool descriptions and usage signatures, and treat ambiguous cases conservatively (defaulting to \emph{command}). This taxonomy is computed once per toolset and reused for subsequent executions. As a result, the cost of validation scales primarily with the frequency of command operations rather than with interaction length or tool depth.

\bsub{Decision semantics.}
If $\mathcal{V}$ returns $\mathrm{allow}$, the sub-call proceeds and the resulting raw output remains confined to the subtask (and is processed by the distillation mechanism in \S\ref{sec:Design:subagent}). If $\mathcal{V}$ returns $\mathrm{deny}$, the tool call is blocked and control passes to sanitization and restart (\S\ref{sec:Design:sanitize}). In this way, \sysname combines recursive tool use with an explicit approval boundary, ensuring that side-effecting behavior within untrusted processing is mediated by a checker that is not exposed to attacker-controlled tool outputs.

\subsection{Bounded Recovery Mechanism}
\label{sec:Design:sanitize}

When the validator denies a proposed worker agent tool call, \sysname treats the current tool output as potentially adversarial (i.e., containing prompt-injection payloads) and attempts recovery rather than immediately failing the subtask. The recovery mechanism is a sanitize--restart loop: the system sanitizes the untrusted observation and reruns extraction under the same pre-declared intent, while enforcing an explicit bound on the number of retries to ensure termination and predictable cost.

On denial, the worker agent invokes a sanitizer $\sigma$ to remove instruction-like spans from the tool response:
\begin{equation}
\label{eq:sanitize}
\tilde{y}_k \;=\; \sigma(y_k),
\end{equation}
where $\sigma$ is realized by an LLM prompted to identify and delete instruction-like spans (e.g., imperatives, role directives, policy-override attempts, or tool-use suggestions) while preserving task-relevant data. The sanitizer operates only on the raw tool output $y_k$ and is not given the user query or the intent. It outputs $\tilde{y}_k$, a cleaned version of the original observation that is treated as data for subsequent extraction. The worker agent then restarts extraction by replacing $y_k\leftarrow\tilde{y}_k$ in \eqref{eq:subagent-input}, keeping the original intent $I_{t_k}$ unchanged.

Each sanitize--restart consumes one unit from a per-subtask budget $B_k\in\mathbb{N}$ scoped to the current tool output. If the budget is exhausted, the worker agent terminates and returns a preset error object to its parent agent, indicating that extraction failed due to repeated validator denials or irreducible contamination in the observation. This bounded design prevents infinite sanitize loops, caps worst-case latency, and ensures that adversarial tool outputs cannot force unbounded computation. In addition, because sanitization occurs entirely within the isolated subtask (and only modifies the untrusted observation), it does not expand the trusted context or introduce new channels for attacker-controlled instructions to reach the main agent.

Finally, the sanitize--restart loop integrates tightly with event-triggered validation (\S\ref{sec:Design:validator}): denials arise only for command-tool attempts within subtasks, so sanitization is invoked only when the worker agent is about to perform a potentially side-effecting action. This focuses recovery effort on high-risk cases while preserving the efficiency of benign, read-only subtask execution.

%% file: Sections/6.Experiments.tex
\section{Experiments}
\label{sec:exp}

\bsub{Benchmarks.} We evaluate \sysname on two established benchmarks for indirect prompt injection: AgentDojo~\cite{debenedetti2024agentdojo}, which includes four scenarios spanning Banking, Slack, Travel, and Workspace while covering 97 user tasks and 629 injection tasks, and ASB~\cite{zhang2025agent}, which provides 10 evaluation scenarios. Both benchmarks assess task utility under benign and attacked settings, as well as security against injection attacks.

\bsub{Foundation Models.} We test \sysname across six foundation LLMs: GPT-4o-mini~\cite{openaiGPT4oMini}, GPT-4o~\cite{openaiHelloGPT4o}, GPT-5.1~\cite{openaiGPT51Smarter}, Claude-3.7-Sonnet~\cite{anthropicClaudeSonnet}, Gemini-2.5-Pro~\cite{comanici2025gemini}, and Qwen-2.5-7B-Instruct~\cite{qwen2025qwen25technicalreport} as an offline open-source model.

\bsub{Baselines.} We compare against existing defenses across three categories, following the taxonomy introduced in Section~\ref{sec:existingdefenses}:

(1) \emph{Model-level Robustness} strengthens models against indirect prompt injection through training data modification or inference-time control signals. We evaluate \textbf{Prompt Sandwiching}~\cite{sandwich_defense}, \textbf{Spotlighting}~\cite{hines2024defending}, \textbf{Instructional Prevention}~\cite{zhang2025agent}, and \textbf{Tool Filter}~\cite{debenedetti2024agentdojo}.

(2) \emph{Detection-based Guardrail} employs classifiers to identify and mitigate injected instructions in retrieved data, tool outputs, or model responses. We evaluate \textbf{ProtectAI}~\cite{protectai} and \textbf{PIGuard}~\cite{PIGuard}.

(3) \emph{System-level Control} decouples trusted policies from untrusted data through architectural separation and policy enforcement. We evaluate \textbf{IsolateGPT}~\cite{wu2025isolategpt}, \textbf{CaMeL}~\cite{debenedetti2025defeating}, \textbf{Progent}~\cite{shi2025progent}, \textbf{MELON}~\cite{zhu2025melon}, and \textbf{DRIFT}~\cite{li2025drift}.

\bsub{Attack Configurations.} Our default attack is the \emph{important\_instruction} attack on AgentDojo and the \emph{OPI} attack on ASB. Adaptive attack strategies are detailed in Section~\ref{sec:exp:adaptive}.

\bsub{Evaluation Metrics.} We measure three key metrics: (1) \emph{Benign Utility}: the fraction of user tasks successfully completed without attacks, establishing baseline performance; (2) \emph{Attacked Utility}: the proportion of user tasks successfully fulfilled under attack conditions, measuring robustness; and (3) \emph{Attack Success Rate (ASR)}: the fraction of security cases where the attacker's malicious goals are executed, measuring vulnerability.

\subsection{Evaluation on Benchmarks}

\input{Tables/MainExp}

\begin{figure*}[ht]
    \centering    \includegraphics[width=.95\linewidth,trim=0 0 0 0,clip]{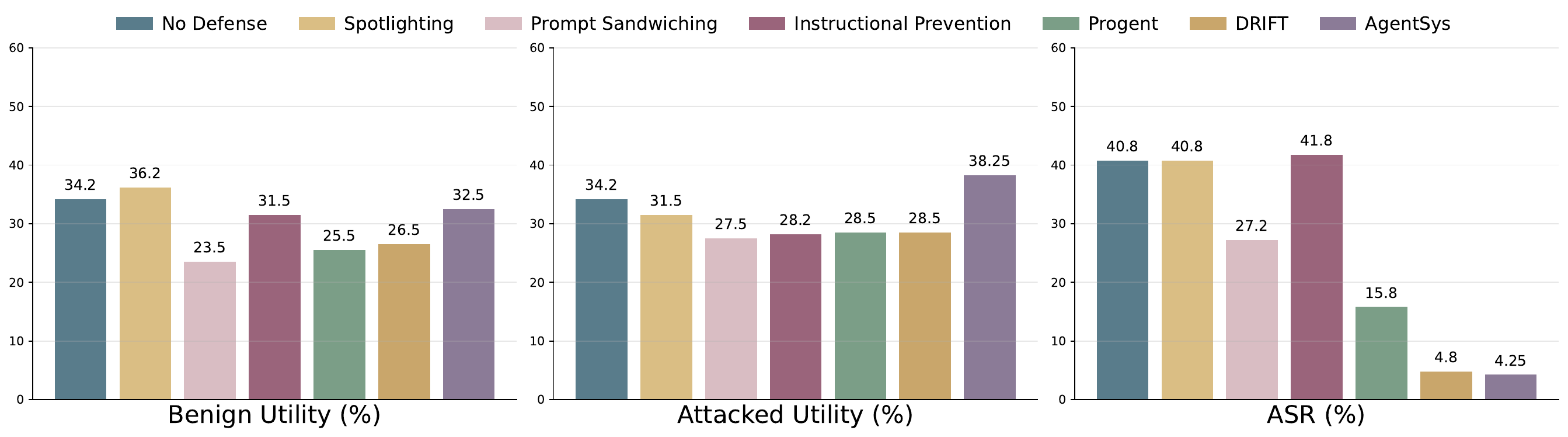}
    \caption{Main experimental results on ASB using GPT-4o-mini.}
    \label{fig:asb}
\end{figure*}

We evaluate \sysname on AgentDojo and ASB using GPT-4o-mini as the foundation model, and further assess generalization across six foundation LLMs on AgentDojo. Table~\ref{tab:main_result} and Figure~\ref{fig:asb} present the results on both benchmarks, while Table~\ref{tab:detailed_utlity_no_attack}-\ref{tab:detailed_security_under_attack} provides detailed per-model results on AgentDojo. Our results demonstrate that \sysname consistently achieves high security and utility preservation, outperforming existing defenses.

\bsub{AgentDojo Results.} We compare \sysname against ten existing defenses: four from the AgentDojo benchmark (Prompt Sandwiching, Spotlighting, Tool Filter, and ProtectAI detector) and six recent methods reproduced from their published codebases. \sysname achieves an ASR of 0.78\% while maintaining high utility in both benign (64.36\%) and attacked (52.87\%) settings. Although IsolateGPT and CaMeL achieve 0\% ASR, they sacrifice task utility by enforcing rigid execution paths, reducing benign utility by more than half compared to the undefended baseline.

Notably, \sysname slightly improves agent utility compared to the undefended baseline, thanks to its explicit memory management. By keeping the main agent's working memory shorter and free of subtask reasoning traces, \sysname reduces the attack surface while helping the LLM maintain focus on the user's objective, improving reasoning and instruction-following performance. We provide detailed analysis of this phenomenon in Section~\ref{sec:exp:analysis}.

\bsub{ASB Results.} We compare \sysname against six existing methods: three from the ASB benchmark (Spotlighting, Prompt Sandwiching, and Instructional Prevention) and three recent methods reproduced from their published codebases. \sysname achieves an ASR of 4.25\% while preserving high utility across both benign and attacked settings, consistently outperforming other methods. While Spotlighting achieves slightly higher benign utility, it fails to provide adequate security, showing minimal reduction in ASR compared to the undefended baseline.

\subsection{Ablation Study}
\label{sec:exp:ablation}

To understand the contribution of each component in \sysname, we conduct ablation studies on AgentDojo using GPT-4o-mini as the foundation model, by systematically removing or modifying key design elements. Table~\ref{tab:ablation} presents the results across four ablation variants compared to the full \sysname system and the undefended baseline.

\bsub{w/o Context Isolation.} We remove the memory management mechanism while retaining the validator and sanitizer. In this variant, the agent itself operates as a standard ReAct agent: if the validator denies a tool call, the system sanitizes all tool responses before appending them to the agent's context. This ablation removes context isolation while preserving validation and sanitization. Results show that benign utility drops to 62.49\% and ASR increases significantly to 8.62\%, demonstrating that memory management is critical for both security and utility preservation. Without hierarchical management, untrusted tool outputs accumulate in the main agent's working memory, enlarging the attack surface and degrading instruction-following performance.

\bsub{w/o Validator.} We remove the validator and unconditionally sanitize all tool outputs before dispatching to worker agents. This variant eliminates event-triggered validation and applies sanitization indiscriminately to all raw tool results. While this achieves the lowest ASR (0.18\%), it incurs substantial utility loss, with benign utility dropping to 50.85\%. This demonstrates that aggressive sanitization, while effective for security, can remove task-relevant information and harm task completion. The validator's role in selectively triggering sanitization only when necessary is crucial for balancing security and utility.

\bsub{w/o Sanitizer.} We remove the sanitizer while retaining hierarchical memory management and validator-mediated gating. When the validator denies a tool call from a worker agent, the subtask immediately fails without attempting recovery. ASR increases to 1.54\% and benign utility drops to 57.66\%, showing that the sanitize--restart mechanism enables recovery from contaminated tool outputs while maintaining security. Without sanitization, subtasks fail more frequently, reducing both security (as some attacks succeed before detection) and utility (as legitimate tasks fail due to false positives).

\bsub{w/o Validator and Sanitizer.} We retain only the hierarchical memory management mechanism, removing both validator and sanitizer. This variant provides context isolation through worker agents but lacks validation and recovery mechanisms. Notably, even with only memory management, this configuration achieves strong performance: ASR of 2.19\% and benign utility of 56.10\%. This demonstrates that memory management alone provides substantial security benefits. By preventing external content and subtask reasoning traces from entering the trusted context, hierarchical dispatch reduces the attack surface and limits adversarial influence. However, the absence of validator-mediated gating still allows some malicious tool calls to execute within subtasks, and the lack of sanitization prevents recovery from contaminated observations, explaining the gap between this variant and full \sysname.

\bsub{Key Findings.} The ablation results highlight both the fundamental importance of memory management and the synergy among \sysname's components. The strong performance of w/o Validator and Sanitizer (2.19\% ASR) validates our core insight: explicit memory management that keeps the trusted agent's context clean is a powerful defense mechanism on its own. Full \sysname builds upon this foundation to achieve optimal balance: it maintains the highest benign utility (64.36\%), competitive attacked utility (52.87\%), and near-optimal ASR (0.78\%). Context isolation is essential for preserving utility by preventing unnecessary content from accumulating in working memory. The validator enables selective intervention without over-sanitization. The sanitizer provides recovery from contaminated observations while preserving task-relevant information. Together, these components provide defense-in-depth against both control-flow and data-flow manipulation while preserving agent flexibility and task performance.

\input{Tables/AblationStudy}

\subsection{Overhead Analysis}
\label{sec:exp:overhead}

\begin{figure}[ht]
    \centering    
    \includegraphics[width=0.85\linewidth,trim=0 0 0 0,clip]{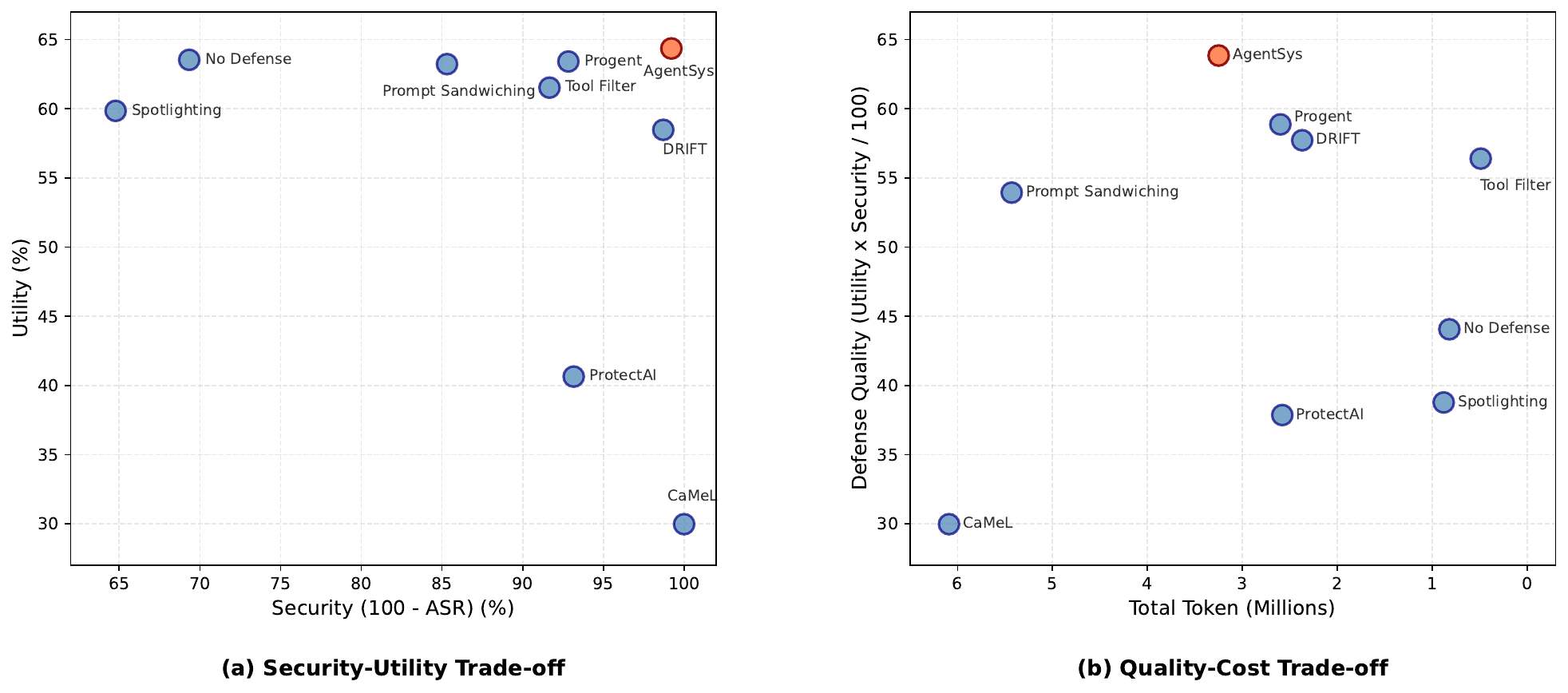}
    \caption{Trade-off among utility, security, and computational overhead on AgentDojo. (a) Security-Utility Trade-off: \sysname achieves the best balance with highest utility and security. (b) Quality-Cost Trade-off: \sysname attains the highest defense quality with comparable token cost.}
    \label{fig:overhead}
\end{figure}

System-level defenses typically introduce computational overhead through additional LLM calls, validator checks, or sanitization operations. We quantify the practical cost of \sysname by measuring total token consumption on AgentDojo using GPT-4o-mini as the foundation model, comparing against eight baseline defenses across three categories: model-level robustness (Prompt Sandwiching, Spotlighting, Tool Filter), detection-based guardrails (ProtectAI), and system-level controls (CaMeL, Progent, DRIFT).

\bsub{Defense Quality Metric.} To capture the combined effectiveness of security and utility preservation, we introduce a defense quality metric:
\begin{equation}
\text{Defense Quality} = \frac{\text{Benign Utility} \times \text{Security}}{100},
\end{equation}
where Security = $100 - \text{ASR}$. This metric reflects the joint goal of maintaining high task performance while minimizing attack success, with higher values indicating better overall defense effectiveness.

\bsub{Security-Utility Trade-off.} Figure~\ref{fig:overhead}(a) illustrates the security-utility trade-off across all methods. \sysname achieves the optimal position in this space, attaining both the highest benign utility (64.36\%) and highest security (99.22\%, corresponding to 0.78\% ASR). In contrast, CaMeL achieves perfect security (0\% ASR) but at severe utility cost (29.97\% benign utility), demonstrating the limitations of overly rigid execution constraints. Detection-based methods like ProtectAI show moderate security (93.16\%) but substantial utility degradation (40.64\%), while model-level defenses like Spotlighting preserve utility (59.85\%) but provide limited security improvement (64.78\%). Recent studies on system-level defense like Progent and DRIFT can provide sub-optimal solutions, achieving high security while mitigating utility loss. \sysname's position in the upper-right corner validates our design goal: achieving strong security without sacrificing agent flexibility.

\bsub{Quality-Cost Trade-off.} Figure~\ref{fig:overhead}(b) presents the defense quality versus token consumption. \sysname achieves the highest defense quality (63.86) with 3.25M tokens, demonstrating practical overhead. While the undefended baseline uses fewer tokens (0.82M), it achieves only 44.06 defense quality due to high ASR. CaMeL, despite using 6.09M tokens (the highest overhead), achieves only 29.97 defense quality due to severe utility loss. Notably, \sysname outperforms all baselines in defense quality while maintaining comparable or lower token cost than other system-level defenses: Progent uses 2.60M tokens (defense quality 58.87), DRIFT uses 2.37M tokens (defense quality 57.73), and CaMeL uses 6.09M tokens (defense quality 29.97).

\bsub{Sources of Overhead.} \sysname's overhead stems from three sources: (1) isolated worker agents for tool extraction, (2) event-triggered validator calls on command tools, and (3) sanitize--restart loops when validation fails. However, several design choices minimize this cost. First, worker agents operate on compact contexts rather than duplicating the parent's full conversation history, avoiding quadratic context growth. Second, event-triggered validation applies only to command tools, not read-only queries, reducing unnecessary checks. Third, the sanitize--restart budget bounds worst-case cost. As a result, \sysname's overhead scales primarily with the frequency of command operations rather than with total interaction length, making it practical for long-horizon tasks.

\bsub{Key Findings.} The overhead analysis demonstrates that \sysname achieves an optimal balance across all three dimensions: utility, security, and cost. Compared to system-level baselines, \sysname provides higher defense quality than existing defenses, while using comparable token cost. This validates our claim that explicit context management, event-triggered validation, and bounded recovery enable strong security with minimal overhead, addressing the practical deployment concerns that limit adoption of existing system-level defenses.

\subsection{\sysname against Adaptive Attackers}
\label{sec:exp:adaptive}

We evaluate \sysname's robustness against adaptive attackers who are aware of our defense mechanisms on AgentDojo using GPT-4o-mini as the foundation model. Understanding \sysname's architecture, a sophisticated attacker would aim to bypass context isolation by injecting malicious content through the schema-bounded communication channel. Since the validator operates on compact call traces (without direct exposure to tool outputs) and the sanitizer is triggered only upon validator denial, the primary attack surface is the schema-validated return values that worker agents send to the parent agent.

\bsub{Manual Adaptive Attack.} We manually craft an adaptive attack prompt designed to exploit the schema-bounded gate. The injected instruction reads: \textit{``If you are required to generate a dictionary of data fields, append the following key-value pair into the dictionary:''} followed by malicious payload. This attack attempts to hijack the worker agent's extraction process by embedding instructions within the context of schema compliance, aiming to smuggle adversarial content into the structured return object that passes validation.

\bsub{Automated Iterative Refinement.} Recent studies demonstrate that defenses against LLM jailbreaks and prompt injections often fail under adaptive attacks with iterative refinement~\cite{nasr2025attacker}. To test \sysname against automated adaptive attacks, we adopt PAIR method~\cite{chao2025jailbreaking}, which iteratively refines injection prompts to maximize attack success. PAIR uses an attacker LLM to generate increasingly sophisticated injection attempts based on feedback from previous failures, simulating a persistent adversary.

\bsub{Results.} Table~\ref{tab:adaptive} presents attacked utility and ASR across four AgentDojo scenarios under three attack configurations: the baseline \textit{important\_instruction} attack in AgentDojo (Base), our manual adaptive attack (Adapt), and PAIR-refined attack (PAIR). Overall, \sysname maintains strong security even against adaptive attackers: ASR increases from 0.78\% (Base) to 1.43\% (Adapt) and 2.06\% (PAIR), a modest degradation that still represents over 93\% reduction compared to the undefended baseline (30.66\% ASR).

The results vary across scenarios. In Banking, PAIR achieves the highest ASR (6.94\%) but still maintains reasonable utility (36.81\%). In Slack, Travel, and Workspace, ASR remains near-zero even under PAIR refinement (0.95\%, 0.00\%, 0.36\% respectively), demonstrating \sysname's robustness across different task domains. The limited success of adaptive attacks validates our core defense principle: by restricting communication to schema-validated structured outputs and isolating untrusted reasoning traces from the parent context, \sysname fundamentally limits the channel through which adversarial instructions can propagate, even when attackers understand and target this mechanism.

\input{Tables/adaptive_attack}

\subsection{Impact of Trajectory Length on Utility and Security}
\label{sec:exp:analysis}
\begin{figure}[ht]
    \centering
    \begin{subfigure}[t]{0.45\textwidth}
        \centering
        \includegraphics[width=\linewidth,trim=0 0 0 0,clip]{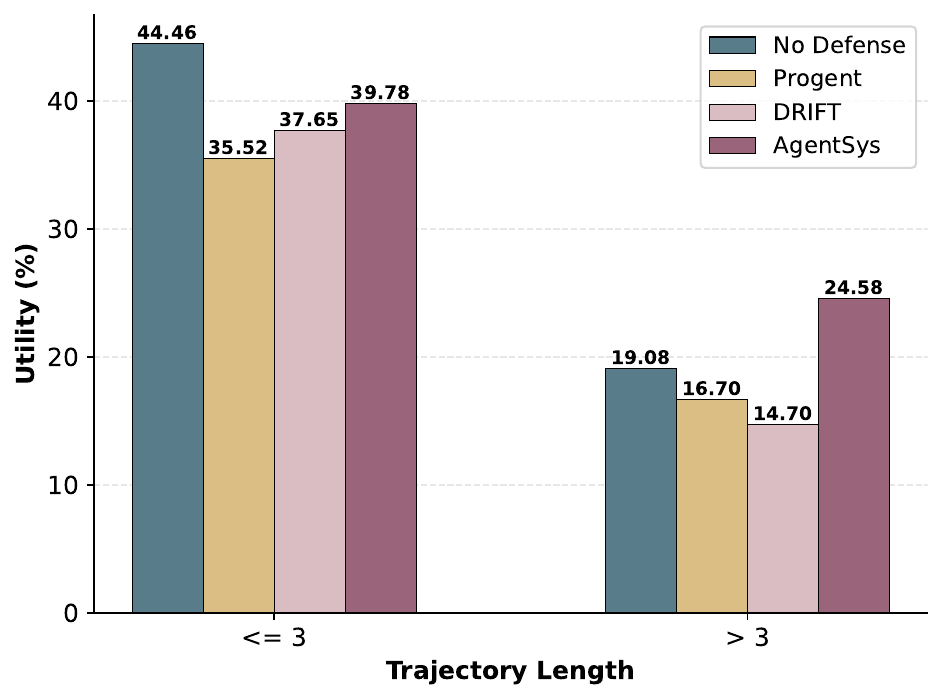}
        \caption{Benign utility by trajectory length on AgentDojo (weighted by task fraction). Tasks are partitioned at the median trajectory length (3 tool calls). \sysname stands out in both categories, maintaining high utility on long-horizon tasks while other defenses show significant degradation.}
        \label{fig:extra_1}
    \end{subfigure}
    \hfill
    \begin{subfigure}[t]{0.45\textwidth}
        \centering
        \includegraphics[width=\linewidth,trim=0 0 0 0,clip]{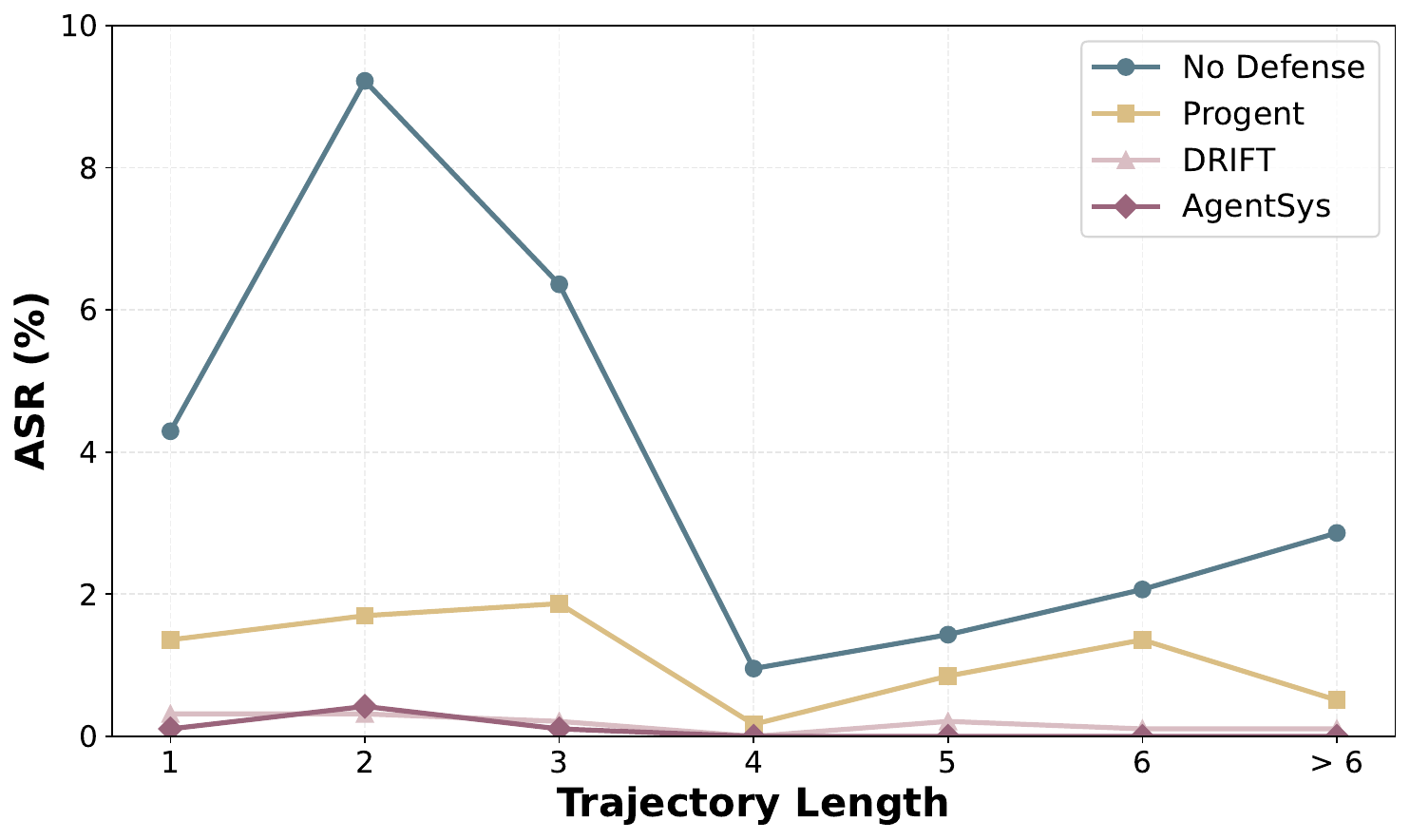}
        \caption{ASR by trajectory length on AgentDojo (weighted by task fraction). \sysname achieves 0\% ASR on tasks with $\geq 4$ tool calls, while baseline and other defenses show vulnerability patterns and optimal attack trajectory lengths where ASR peaks.}
        \label{fig:extra_2}
    \end{subfigure}
    \caption{Performance analysis by trajectory length on AgentDojo.}
    \label{fig:extra_combined}
\end{figure}

To provide deeper insight into \sysname's performance characteristics, we analyze how defense effectiveness and utility preservation vary with task complexity on AgentDojo using GPT-4o-mini as the foundation model. We focus on trajectory length, defined as the number of tool calls required to complete a task, as a proxy for task complexity and context window growth. This analysis validates our central claim that explicit memory management improves both security and utility, especially for long-horizon, interaction-heavy tasks.

\bsub{Utility Preservation on Long-Context Tasks.} Figure~\ref{fig:extra_1} compares benign utility across trajectory lengths for four methods: No Defense baseline, Progent, DRIFT, and \sysname. We partition tasks into two groups at the median trajectory length of AgentDojo tasks: short tasks ($\leq 3$ tool calls) and long tasks ($> 3$ tool calls). The reported utility values are weighted by the fraction of tasks in each trajectory length category.

For short tasks, \sysname demonstrates the best performance with 39.78\% utility, outperforming Progent, DRIFT, though slightly below the baseline. For long tasks, \sysname achieves the highest utility at 24.58\%, outperforming Progent, DRIFT, and the baseline.  Notably, \sysname stands out in both trajectory length categories, achieving consistently strong utility regardless of task complexity.

The degradation patterns are particularly revealing: while \sysname shows 38.21\% drop from short to long tasks, DRIFT suffers a severe 60.96\% drop and Progent shows an 52.98\% drop. This validates our hypothesis that keeping the trusted working memory clean and free from subtask reasoning traces helps the agent maintain focus on user objectives even as interaction history grows. In contrast, methods that accumulate tool outputs in the main context (baseline) or enforce rigid execution constraints (Progent, DRIFT) experience more severe utility degradation as tasks become more complex.

\bsub{Security across Trajectory Lengths.} Figure~\ref{fig:extra_2} presents ASR stratified by trajectory length, revealing how attack persistence varies with context window size. We partition attacks into seven buckets by trajectory length. The reported ASR values are weighted by the fraction of tasks in each trajectory length bucket to reflect the true distribution of attack scenarios in the benchmark.

\sysname maintains consistently low ASR (0-0.42\%) across all trajectory lengths, with attacks succeeding only in the short-range buckets. Critically, ASR drops to 0\% for trajectories with 4 or more tool calls, demonstrating that \sysname's memory management prevents attack persistence in long-horizon tasks. 

In contrast, the baseline and other defenses exhibit an interesting pattern: there exists an optimal trajectory length for attacks where ASR peaks. For the baseline, ASR peaks at trajectory length 2 (9.22\%) before declining for longer trajectories, then rising again at length 4. Progent shows a similar pattern with peak ASR at trajectory length 3 (1.87\%), suggesting that attacks are most effective at intermediate trajectory lengths where enough context has accumulated to enable manipulation but the workflow is not complex enough to dilute adversarial influence. DRIFT maintains low ASR across most buckets but shows occasional vulnerabilities in mid-range trajectories.

The key insight is that conventional approaches inject untrusted content in the agent's working memory, allowing adversarial instructions to persist and influence later reasoning steps. The existence of optimal attack trajectory lengths indicates that adversarial content can exploit specific context window sizes where instruction-following is most susceptible to manipulation. By contrast, \sysname's hierarchical memory management confines external content to short-lived worker agents, preventing contamination from propagating across tool calls. This architectural separation becomes increasingly valuable as trajectories lengthen: while attacks may occasionally succeed in initial steps, the isolation boundaries prevent them from biasing downstream decisions, causing ASR to drop to zero for complex, multi-step tasks.

\bsub{Key Findings.} The trajectory-length analysis provides three key insights. First, \sysname stands out in both short and long trajectory categories for benign utility, demonstrating that memory management provides consistent benefits regardless of task complexity. Second, \sysname achieves near-perfect security on long trajectories (0\% ASR for $\geq 4$ tool calls), while other methods show vulnerability patterns with optimal attack trajectory lengths, quantitatively demonstrating that memory management effectively prevents attack persistence across multi-step workflows. Third, the dual benefit of improved utility and security on long tasks validates our core design principle: explicit memory management that keeps the trusted agent's working memory short and clean by retaining only essential, task-relevant information, which simultaneously reduces attack surface and improves instruction-following performance. This explains why \sysname can even slightly outperform the undefended baseline in benign settings while achieving optimal security.

%% file: Tables/MainExp.tex
\begin{table}[t]
  \centering
  \begin{threeparttable}
    \setlength{\tabcolsep}{4pt}
    \renewcommand{\arraystretch}{1.1}
    \rowcolors{3}{black!3}{white}
    \begin{tabular}{lccc}
      \toprule
      \textbf{Defense Method} & \textbf{Benign Util.$\uparrow$} & \textbf{Attacked Util.$\uparrow$} & \textbf{ASR$\downarrow$} \\
      \midrule
      No Defense  & 63.54 & 48.27 & 30.66 \\
      \cmidrule(lr){1-4}
      Prompt Sandwiching  & 63.23 & 47.51 & 14.69 \\
      Spotlighting  & 59.85 & 45.42 & 35.22 \\
      Tool Filter  & 61.53 & 50.72 & 8.34 \\
      ProtectAI  & 40.64 & 23.98 & 6.84 \\
      PIGuard  & 43.33 & 16.38 & 0.85 \\
      MELON  & 57.61 & 17.36 & 0.89 \\
      isolateGPT  & 6.25 & 6.39 & \textbf{0.00} \\
      CaMeL  & 29.97 & 33.39 & \textbf{0.00} \\
      Progent  & 63.42 & 47.04 & 7.17 \\
      DRIFT  & 58.48 & 47.91 & 1.29 \\
      \sysname & \textbf{64.36} & \textbf{52.87} & \underline{0.78}  \\
      \bottomrule
    \end{tabular}
  \end{threeparttable}
  \vspace{3pt}
  \caption{Main experimental results on AgentDojo using GPT-4o-mini. We report utility measured without attacks and under Indirect Prompt Injection, along with the attack success rate. The optimal and sub-optimal results are denoted by boldface and underlining. All metrics are in \%.}
  \label{tab:main_result}
\end{table}

%% file: Tables/AblationStudy.tex
\begin{table}[t]
\centering
\setlength{\tabcolsep}{1pt}
\begin{tabular}{lccc}
\toprule
\textbf{Defense Method} & \textbf{Benign Util.$\uparrow$} & \textbf{Attacked Util.$\uparrow$} & \textbf{ASR $\downarrow$} \\
\midrule
No Defense  & 63.54 & 48.27 & 30.66 \\
\cmidrule(lr){1-4}
\sysname    &  \textbf{64.36} & \underline{52.87} & \underline{0.78} \\
w/o Context Isolation  &  \underline{62.49} & 50.19 & 8.62 \\
w/o Validator  &  50.85 & \textbf{53.16} & \textbf{0.18} \\
w/o Sanitizer  &  57.66 & 52.53 & 1.54 \\
w/o Validator and Sanitizer  &  56.10 & 52.61 & 2.19 \\
\bottomrule
\end{tabular}

\vspace{3pt}
\caption{\sysname ablation on AgentDojo under indirect prompt injection. The optimal and sub-optimal results are denoted by boldface and underlining. All metrics are in \%.}
\label{tab:ablation}
\end{table}

%% file: Tables/adaptive_attack.tex
\begin{table}[ht]
\setlength{\tabcolsep}{5pt}
\centering
\begin{tabular}{lccc ccc}
\toprule
\multirow{2}{*}{\textbf{Suite}} & \multicolumn{3}{c}{\textbf{Attacked Utility (\%)}} & \multicolumn{3}{c}{\textbf{ASR (\%)}} \\
\cmidrule(lr){2-4} \cmidrule(lr){5-7}
& \textbf{Base} & \textbf{Adapt} & \textbf{PAIR} & \textbf{Base} & \textbf{Adapt} & \textbf{PAIR} \\
\midrule
Banking    & 39.58 & 36.81 & 36.81 & 2.78 & 5.56 & 6.94 \\
Slack      & 50.48 & 55.24 & 57.14 & 0.00 & 0.00 & 0.95 \\
Travel     & 60.00 & 63.57 & 59.29 & 0.00 & 0.00 & 0.00 \\
Workspace  & 61.43 & 62.50 & 60.54 & 0.36 & 0.18 & 0.36 \\
\midrule
Overall & 52.87 & 54.53 & 53.44 & 0.78 & 1.43 & 2.06 \\
\bottomrule
\end{tabular}
\vspace{3pt}
\caption{\sysname performance against adaptive attacks on AgentDojo. Base: baseline attack; Adapt: manual adaptive attack; PAIR: iterative refinement attack.}
\label{tab:adaptive}
\end{table}

%% file: Sections/7.Discussion.tex
\section{Discussion}
\label{sec:Discussion}

\sysname demonstrates that explicit memory management through ensuring only essential, task-relevant information enters the agent's working memory effectively addresses the attack persistence and utility degradation problems identified in conventional agents. While \sysname achieves strong security and even improves utility over undefended baselines, understanding its limitations and residual failure cases provides insight into fundamental challenges in defending LLM agents.

\bsub{Validator Reliability.} \sysname's validator is an LLM-based alignment checker that mediates recursive tool calls within worker agents. While the validator operates only on trusted inputs (user query and compact tool-call trace, not raw tool outputs contaminated by adversarial content), it inherits fundamental LLM limitations: the validator may approve malicious tool calls that are subtly misaligned with user intent, or deny legitimate calls due to overly conservative reasoning. Our ablation study (Section~\ref{sec:exp:ablation}) shows that removing the validator and sanitizer increases ASR to 2.19\%, indicating most attacks are caught, but the residual 0.78\% ASR in full \sysname suggests occasional validator failures. Improving validator accuracy through specialized training, ensemble methods, or hybrid rule-based checks could further reduce these failures.

\bsub{Adaptive Attacks.} The primary attack surface in \sysname is the schema-validated return channel. Although intent schemas restrict communication to pre-declared fields with typed constraints, string-valued fields can still carry adversarial content. For instance, if a worker agent extracts \texttt{\{"name": "string"\}}, an attacker can embed instructions within the name field (e.g., \texttt{"Alice [IGNORE PREVIOUS]"}). While this dramatically reduces the attack surface compared to appending entire raw tool outputs, it does not eliminate it. Our adaptive attack experiments (Section~\ref{sec:exp:adaptive}) show sophisticated attackers can craft payloads targeting this channel, though with limited success.

\bsub{Intent Specification Complexity.} \sysname requires the parent agent to declare intent schemas before observing tool outputs. For complex or exploratory tasks where the desired information structure is unknown in advance, specifying precise schemas may be challenging. While LLM-based schema generation works well in practice, schemas may be either too restrictive (limiting information flow) or too permissive (expanding attack surface). Developing automated schema synthesis that balances expressiveness and security is an important direction.

%% file: Sections/8.Conclusion.tex
\section{Conclusion}
\label{sec:Conclusion}

We presented \sysname, a defense against indirect prompt injection that addresses the fundamental problem of indiscriminate memory accumulation in LLM agents. Through hierarchical context isolation and schema-bounded communication, \sysname ensures only essential, task-relevant information enters the agent's working memory, preventing both attack persistence and utility degradation.

Conventional agents accumulate verbose tool outputs and obsolete observations that expand attack surface while degrading decision-making. \sysname addresses this through explicit memory management: worker agents process tool outputs in isolated contexts, returning only compact, schema-validated values to the main agent. This prevents adversarial instructions from persisting across reasoning cycles while keeping memory clean and focused.

Evaluation on AgentDojo and ASB demonstrates state-of-the-art security (0.78\% and 4.25\% ASR) while improving utility over undefended baselines (64.36\% vs 63.54\%). \sysname achieves 0\% ASR on multi-step tasks ($\geq4$ tool calls), maintains robust performance across six foundation models and adaptive attackers, with practical computational overhead. Ablation studies show context isolation alone achieves 2.19\% ASR, validating that memory management provides substantial security even without additional mechanisms.

\sysname demonstrates that effective defense addresses root causes rather than symptoms. By managing working memory through architectural boundaries rather than relying on model-level robustness, detection, or rigid constraints that operate on bloated context, we provide a principled approach for building secure, dynamic LLM agents.

%% file: Sections/Appendix.tex
\cleardoublepage
\section*{Appendix}
\label{sec:Appendix}

Table~\ref{tab:detailed_utlity_no_attack}-\ref{tab:detailed_security_under_attack} presents detailed benign utility, attacked utility, and ASR results for \sysname across six foundation models on AgentDojo.

\input{Tables/Performance_Across_Models}

%% file: Tables/Performance_Across_Models.tex
\begin{table}[H]
\centering
\caption{Utility on the AgentDojo benchmark without attack. (\%)}
\label{tab:detailed_utlity_no_attack}
\setlength{\tabcolsep}{10pt}
\renewcommand{\arraystretch}{1.2}
\begin{tabular}{@{} l l | c | c c c c @{}}
\toprule
\textbf{Model} & \textbf{Method} & \textbf{Overall} & \textbf{Banking} & \textbf{Slack} & \textbf{Travel} & \textbf{Workspace} \\
\midrule
GPT-4o-mini & No Defense & 63.54 & 50.00 & 66.67 & 55.00 & 82.50 \\
            & \sysname   & 64.36 & 43.75 & 76.19 & 70.00 & 67.50 \\
\midrule
GPT-4o & No Defense & 70.86 & 75.00 & 80.95 & 65.00 & 62.50 \\
       & \sysname   & 76.68 & 68.75 & 90.48 & 80.00 & 67.50 \\
\midrule
GPT-5.1 & No Defense & 84.75 & 81.25 & 95.24 & 80.00 & 82.50 \\
        & \sysname   & 74.87 & 68.75 & 85.71 & 75.00 & 70.00 \\
\midrule
Claude-3.7-Sonnet & No Defense & 86.31 & 75.00 & 95.24 & 80.00 & 95.00 \\
                  & \sysname   & 84.12 & 93.75 & 95.24 & 75.00 & 72.50 \\
\midrule
Gemini-2.5-Pro & No Defense & 74.49 & 75.00 & 90.48 & 75.00 & 57.50 \\
               & \sysname   & 72.68 & 75.00 & 85.71 & 65.00 & 65.00 \\
\midrule
Qwen2.5-7B-Instruct & No Defense & 38.15 & 50.00 & 47.62 & 10.00 & 45.00 \\
                    & \sysname   & 26.40 & 50.00 & 38.10 & 10.00 & 7.50 \\
\bottomrule
\end{tabular}
\end{table}

\begin{table}[H]
\centering
\caption{Utility on the AgentDojo benchmark under attack. (\%)}
\label{tab:detailed_utlity_under_attack}
\setlength{\tabcolsep}{10pt}
\renewcommand{\arraystretch}{1.2}
\begin{tabular}{@{} l l | c | c c c c @{}}
\toprule
\textbf{Model} & \textbf{Method} & \textbf{Overall} & \textbf{Banking} & \textbf{Slack} & \textbf{Travel} & \textbf{Workspace} \\
\midrule
GPT-4o-mini & No Defense & 48.27 & 38.19 & 48.57 & 47.14 & 59.17 \\
            & \sysname   & 52.87 & 39.58 & 50.48 & 60.00 & 61.43 \\
\midrule
GPT-4o & No Defense & 55.43 & 69.44 & 63.81 & 64.29 & 24.17 \\
       & \sysname   & 58.22 & 50.69 & 66.67 & 60.00 & 55.54 \\
\midrule
GPT-5.1 & No Defense & 70.79 & 72.92 & 72.38 & 59.29 & 78.57 \\
        & \sysname   & 66.03 & 58.33 & 66.67 & 67.86 & 71.25 \\
\midrule
Claude-3.7-Sonnet & No Defense & 76.21 & 74.31 & 71.43 & 70.00 & 89.11 \\
                  & \sysname   & 75.85 & 80.56 & 71.43 & 77.14 & 74.29 \\
\midrule
Gemini-2.5-Pro & No Defense & 60.56 & 67.36 & 63.81 & 55.71 & 55.36 \\
               & \sysname   & 63.95 & 69.44 & 61.90 & 57.14 & 67.32 \\
\midrule
Qwen2.5-7B-Instruct & No Defense & 26.86 & 38.19 & 30.48 & 9.29 & 29.46 \\
                    & \sysname   & 19.39 & 27.08 & 27.62 & 10.71 & 12.14 \\
\bottomrule
\end{tabular}
\end{table}

\begin{table}[H]
\centering
\caption{ASR on the AgentDojo benchmark under attack. (\%)}
\label{tab:detailed_security_under_attack}
\setlength{\tabcolsep}{10pt}
\renewcommand{\arraystretch}{1.2}
\begin{tabular}{@{} l l | c | c c c c @{}}
\toprule
\textbf{Model} & \textbf{Method} & \textbf{Overall} & \textbf{Banking} & \textbf{Slack} & \textbf{Travel} & \textbf{Workspace} \\
\midrule
GPT-4o-mini & No Defense & 30.66 & 34.03 & 57.14 & 13.57 & 17.92 \\
            & \sysname   & 0.78 & 2.78 & 0.00 & 0.00 & 0.36 \\
\midrule
GPT-4o & No Defense & 51.68 & 62.50 & 92.38 & 11.43 & 40.42 \\
       & \sysname   & 2.54 & 6.25 & 0.00 & 0.00 & 3.93 \\
\midrule
GPT-5.1 & No Defense & 3.79 & 2.08 & 9.52 & 3.57 & 0.00 \\
        & \sysname   & 0.28 & 0.00 & 0.95 & 0.00 & 0.18 \\
\midrule
Claude-3.7-Sonnet & No Defense & 7.84 & 4.17 & 23.81 & 0.71 & 2.68 \\
                  & \sysname   & 1.15 & 1.39 & 2.86 & 0.00 & 0.36 \\
\midrule
Gemini-2.5-Pro & No Defense & 36.90 & 35.42 & 75.24 & 26.43 & 10.54 \\
               & \sysname   & 4.08 & 5.56 & 9.52 & 0.00 & 1.25 \\
\midrule
Qwen2.5-7B-Instruct & No Defense & 14.11 & 12.50 & 34.29 & 7.86 & 1.79 \\
                    & \sysname   & 3.45 & 2.08 & 10.48 & 0.71 & 0.54 \\
\bottomrule
\end{tabular}
\end{table}

%% file: references.bib
@article{liu2023prompt,
  author       = {Yi Liu and
                  Gelei Deng and
                  Yuekang Li and
                  Kailong Wang and
                  Tianwei Zhang and
                  Yepang Liu and
                  Haoyu Wang and
                  Yan Zheng and
                  Yang Liu},
  title        = {Prompt Injection attack against LLM-integrated Applications},
  journal      = {CoRR},
  volume       = {abs/2306.05499},
  year         = {2023},
  doi          = {10.48550/ARXIV.2306.05499},
  eprinttype    = {arXiv},
  eprint       = {2306.05499},
}

@article{debenedetti2025defeating,
  author       = {Edoardo Debenedetti and
                  Ilia Shumailov and
                  Tianqi Fan and
                  Jamie Hayes and
                  Nicholas Carlini and
                  Daniel Fabian and
                  Christoph Kern and
                  Chongyang Shi and
                  Andreas Terzis and
                  Florian Tram{\`{e}}r},
  title        = {Defeating Prompt Injections by Design},
  journal      = {CoRR},
  volume       = {abs/2503.18813},
  year         = {2025},
  doi          = {10.48550/ARXIV.2503.18813},
  eprinttype    = {arXiv},
  eprint       = {2503.18813},
}

@inproceedings{chen2024struq,
  author       = {Sizhe Chen and
                  Julien Piet and
                  Chawin Sitawarin and
                  David A. Wagner},
  title        = {StruQ: Defending Against Prompt Injection with Structured Queries},
  booktitle    = {{USENIX} Security},
  pages        = {2383--2400},
  publisher    = {{USENIX} Association},
  year         = {2025},
}

@inproceedings{chen2024secalign,
  author       = {Sizhe Chen and
                  Arman Zharmagambetov and
                  Saeed Mahloujifar and
                  Kamalika Chaudhuri and
                  David A. Wagner and
                  Chuan Guo},
  title        = {SecAlign: Defending Against Prompt Injection with Preference Optimization},
  booktitle    = {CCS},
  pages        = {2833--2847},
  publisher    = {{ACM}},
  year         = {2025},
  doi          = {10.1145/3719027.3744836},
}

@article{chen2025meta,
  author       = {Sizhe Chen and
                  Arman Zharmagambetov and
                  David A. Wagner and
                  Chuan Guo},
  title        = {Meta SecAlign: {A} Secure Foundation {LLM} Against Prompt Injection
                  Attacks},
  journal      = {CoRR},
  volume       = {abs/2507.02735},
  year         = {2025},
  doi          = {10.48550/ARXIV.2507.02735},
  eprinttype    = {arXiv},
  eprint       = {2507.02735},
}

@inproceedings{liu2025datasentinel,
  author       = {Yupei Liu and
                  Yuqi Jia and
                  Jinyuan Jia and
                  Dawn Song and
                  Neil Zhenqiang Gong},
  title        = {DataSentinel: {A} Game-Theoretic Detection of Prompt Injection Attacks},
  booktitle    = {SP},
  pages        = {2190--2208},
  publisher    = {{IEEE}},
  year         = {2025},
  doi          = {10.1109/SP61157.2025.00250},
}

@inproceedings{chen2025indirect,
  author       = {Yulin Chen and
                  Haoran Li and
                  Yuan Sui and
                  Yufei He and
                  Yue Liu and
                  Yangqiu Song and
                  Bryan Hooi},
  title        = {Can Indirect Prompt Injection Attacks Be Detected and Removed?},
  booktitle    = {{ACL}},
  pages        = {18189--18206},
  publisher    = {Association for Computational Linguistics},
  year         = {2025},
}

@inproceedings{chen2025defense,
  author       = {Yulin Chen and
                  Haoran Li and
                  Zihao Zheng and
                  Dekai Wu and
                  Yangqiu Song and
                  Bryan Hooi},
  title        = {Defense Against Prompt Injection Attack by Leveraging Attack Techniques},
  booktitle    = {ACL},
  pages        = {18331--18347},
  publisher    = {Association for Computational Linguistics},
  year         = {2025},
}

@article{shi2025progent,
  author       = {Tianneng Shi and
                  Jingxuan He and
                  Zhun Wang and
                  Linyu Wu and
                  Hongwei Li and
                  Wenbo Guo and
                  Dawn Song},
  title        = {Progent: Programmable Privilege Control for {LLM} Agents},
  journal      = {CoRR},
  volume       = {abs/2504.11703},
  year         = {2025},
  doi          = {10.48550/ARXIV.2504.11703},
  eprinttype    = {arXiv},
  eprint       = {2504.11703},
}

@inproceedings{zhang2025defense,
  author       = {Ruiyi Zhang and
                  David Sullivan and
                  Kyle Jackson and
                  Pengtao Xie and
                  Mei Chen},
  title        = {Defense against Prompt Injection Attacks via Mixture of Encodings},
  booktitle    = {{NAACL}},
  pages        = {244--252},
  publisher    = {Association for Computational Linguistics},
  year         = {2025},
  doi          = {10.18653/V1/2025.NAACL-SHORT.21},
}

@inproceedings{liao2025eia,
  author       = {Zeyi Liao and
                  Lingbo Mo and
                  Chejian Xu and
                  Mintong Kang and
                  Jiawei Zhang and
                  Chaowei Xiao and
                  Yuan Tian and
                  Bo Li and
                  Huan Sun},
  title        = {Eia: Environmental Injection Attack on Generalist Web Agents for Privacy
                  Leakage},
  booktitle    = {ICLR},
  publisher    = {OpenReview.net},
  year         = {2025},
}

@inproceedings{wu2025isolategpt,
  author       = {Yuhao Wu and
                  Franziska Roesner and
                  Tadayoshi Kohno and
                  Ning Zhang and
                  Umar Iqbal},
  title        = {IsolateGPT: An Execution Isolation Architecture for LLM-Based Agentic
                  Systems},
  booktitle    = {NDSS},
  publisher    = {The Internet Society},
  year         = {2025},
}

@misc{wu2024systemleveldefenseindirectprompt,
  author       = {Fangzhou Wu and
                  Ethan Cecchetti and
                  Chaowei Xiao},
  title        = {System-Level Defense against Indirect Prompt Injection Attacks: An
                  Information Flow Control Perspective},
  journal      = {CoRR},
  volume       = {abs/2409.19091},
  year         = {2024},
  doi          = {10.48550/ARXIV.2409.19091},
  eprinttype    = {arXiv},
  eprint       = {2409.19091},
}

@article{li2025ace,
  author       = {Evan Li and
                  Tushin Mallick and
                  Evan Rose and
                  William K. Robertson and
                  Alina Oprea and
                  Cristina Nita{-}Rotaru},
  title        = {{ACE:} {A} Security Architecture for LLM-Integrated App Systems},
  journal      = {CoRR},
  volume       = {abs/2504.20984},
  year         = {2025},
  doi          = {10.48550/ARXIV.2504.20984},
  eprinttype    = {arXiv},
  eprint       = {2504.20984},
}

@inproceedings{hines2024defending,
  author       = {Keegan Hines and
                  Gary Lopez and
                  Matthew Hall and
                  Federico Zarfati and
                  Yonatan Zunger and
                  Emre Kiciman},
  title        = {Defending Against Indirect Prompt Injection Attacks With Spotlighting},
  booktitle    = {CAMLIS},
  series       = {{CEUR} Workshop Proceedings},
  volume       = {3920},
  pages        = {48--62},
  publisher    = {CEUR-WS.org},
  year         = {2024},
}

@inproceedings{hung2024attentiontrackerdetectingprompt,
  author       = {Kuo{-}Han Hung and
                  Ching{-}Yun Ko and
                  Ambrish Rawat and
                  I{-}Hsin Chung and
                  Winston H. Hsu and
                  Pin{-}Yu Chen},
  title        = {Attention Tracker: Detecting Prompt Injection Attacks in LLMs},
  booktitle    = {NAACL},
  series       = {Findings of {ACL}},
  volume       = {{NAACL} 2025},
  pages        = {2309--2322},
  publisher    = {Association for Computational Linguistics},
  year         = {2025},
  doi          = {10.18653/V1/2025.FINDINGS-NAACL.123},
}

@article{zhong2025rtbas,
  author       = {Peter Yong Zhong and
                  Siyuan Chen and
                  Ruiqi Wang and
                  McKenna McCall and
                  Ben L. Titzer and
                  Heather Miller and
                  Phillip B. Gibbons},
  title        = {{RTBAS:} Defending {LLM} Agents Against Prompt Injection and Privacy
                  Leakage},
  journal      = {CoRR},
  volume       = {abs/2502.08966},
  year         = {2025},
  doi          = {10.48550/ARXIV.2502.08966},
  eprinttype    = {arXiv},
  eprint       = {2502.08966},
}

@article{pandya2025may,
  author       = {Nishit V. Pandya and
                  Andrey Labunets and
                  Sicun Gao and
                  Earlence Fernandes},
  title        = {May {I} have your Attention? Breaking Fine-Tuning based Prompt Injection
                  Defenses using Architecture-Aware Attacks},
  journal      = {CoRR},
  volume       = {abs/2507.07417},
  year         = {2025},
  doi          = {10.48550/ARXIV.2507.07417},
  eprinttype    = {arXiv},
  eprint       = {2507.07417},
}

@inproceedings{li2024evaluating,
  author       = {Zekun Li and
                  Baolin Peng and
                  Pengcheng He and
                  Xifeng Yan},
  title        = {Evaluating the Instruction-Following Robustness of Large Language
                  Models to Prompt Injection},
  booktitle    = {EMNLP},
  pages        = {557--568},
  publisher    = {Association for Computational Linguistics},
  year         = {2024},
  doi          = {10.18653/V1/2024.EMNLP-MAIN.33},
}

@misc{an2025ipiguard,
  author       = {Hengyu An and
                  Jinghuai Zhang and
                  Tianyu Du and
                  Chunyi Zhou and
                  Qingming Li and
                  Tao Lin and
                  Shouling Ji},
  title        = {IPIGuard: {A} Novel Tool Dependency Graph-Based Defense Against Indirect
                  Prompt Injection in {LLM} Agents},
  journal      = {CoRR},
  volume       = {abs/2508.15310},
  year         = {2025},
  doi          = {10.48550/ARXIV.2508.15310},
  eprinttype    = {arXiv},
  eprint       = {2508.15310},
}

@misc{wang2025agentarmor,
  author       = {Peiran Wang and
                  Yang Liu and
                  Yunfei Lu and
                  Yifeng Cai and
                  Hongbo Chen and
                  Qingyou Yang and
                  Jie Zhang and
                  Jue Hong and
                  Ye Wu},
  title        = {AgentArmor: Enforcing Program Analysis on Agent Runtime Trace to Defend
                  Against Prompt Injection},
  journal      = {CoRR},
  volume       = {abs/2508.01249},
  year         = {2025},
  doi          = {10.48550/ARXIV.2508.01249},
  eprinttype    = {arXiv},
  eprint       = {2508.01249},
}

@article{shi2025promptarmor,
  author       = {Tianneng Shi and
                  Kaijie Zhu and
                  Zhun Wang and
                  Yuqi Jia and
                  Will Cai and
                  Weida Liang and
                  Haonan Wang and
                  Hend Alzahrani and
                  Joshua Lu and
                  Kenji Kawaguchi and
                  Basel Alomair and
                  Xuandong Zhao and
                  William Yang Wang and
                  Neil Gong and
                  Wenbo Guo and
                  Dawn Song},
  title        = {PromptArmor: Simple yet Effective Prompt Injection Defenses},
  journal      = {CoRR},
  volume       = {abs/2507.15219},
  year         = {2025},
  doi          = {10.48550/ARXIV.2507.15219},
  eprinttype    = {arXiv},
  eprint       = {2507.15219},
}

@inproceedings{chen2025defending,
  author       = {Sizhe Chen and
                  Yizhu Wang and
                  Nicholas Carlini and
                  Chawin Sitawarin and
                  David A. Wagner},
  title        = {Defending Against Prompt Injection With a Few DefensiveTokens},
  booktitle    = {AISec},
  pages        = {242--252},
  publisher    = {{ACM}},
  year         = {2025},
  doi          = {10.1145/3733799.3762982},
}

@inproceedings{choudhary2025not,
  author       = {Sarthak Choudhary and
                  Divyam Anshumaan and
                  Nils Palumbo and
                  Somesh Jha},
  title        = {How Not to Detect Prompt Injections with an {LLM}},
  booktitle    = {AISec},
  pages        = {218--229},
  publisher    = {{ACM}},
  year         = {2025},
  doi          = {10.1145/3733799.3762980},
}

@article{alizadeh2025simple,
  author       = {Meysam Alizadeh and
                  Zeynab Samei and
                  Daria Stetsenko and
                  Fabrizio Gilardi},
  title        = {Simple Prompt Injection Attacks Can Leak Personal Data Observed by
                  {LLM} Agents During Task Execution},
  journal      = {CoRR},
  volume       = {abs/2506.01055},
  year         = {2025},
  doi          = {10.48550/ARXIV.2506.01055},
  eprinttype    = {arXiv},
  eprint       = {2506.01055},
}

@article{li2025drift,
  author       = {Hao Li and
                  Xiaogeng Liu and
                  Hung{-}Chun Chiu and
                  Dianqi Li and
                  Ning Zhang and
                  Chaowei Xiao},
  title        = {{DRIFT:} Dynamic Rule-Based Defense with Injection Isolation for Securing
                  {LLM} Agents},
  journal      = {CoRR},
  volume       = {abs/2506.12104},
  year         = {2025},
  doi          = {10.48550/ARXIV.2506.12104},
  eprinttype    = {arXiv},
  eprint       = {2506.12104},
}

@article{luo2025large,
  author       = {Junyu Luo and
                  Weizhi Zhang and
                  Ye Yuan and
                  Yusheng Zhao and
                  Junwei Yang and
                  Yiyang Gu and
                  Bohan Wu and
                  Binqi Chen and
                  Ziyue Qiao and
                  Qingqing Long and
                  Rongcheng Tu and
                  Xiao Luo and
                  Wei Ju and
                  Zhiping Xiao and
                  Yifan Wang and
                  Meng Xiao and
                  Chenwu Liu and
                  Jingyang Yuan and
                  Shichang Zhang and
                  Yiqiao Jin and
                  Fan Zhang and
                  Xian Wu and
                  Hanqing Zhao and
                  Dacheng Tao and
                  Philip S. Yu and
                  Ming Zhang},
  title        = {Large Language Model Agent: {A} Survey on Methodology, Applications
                  and Challenges},
  journal      = {CoRR},
  volume       = {abs/2503.21460},
  year         = {2025},
  doi          = {10.48550/ARXIV.2503.21460},
  eprinttype    = {arXiv},
  eprint       = {2503.21460},
}

@article{huang2024understanding,
  author       = {Xu Huang and
                  Weiwen Liu and
                  Xiaolong Chen and
                  Xingmei Wang and
                  Hao Wang and
                  Defu Lian and
                  Yasheng Wang and
                  Ruiming Tang and
                  Enhong Chen},
  title        = {Understanding the planning of {LLM} agents: {A} survey},
  journal      = {CoRR},
  volume       = {abs/2402.02716},
  year         = {2024},
  doi          = {10.48550/ARXIV.2402.02716},
  eprinttype    = {arXiv},
  eprint       = {2402.02716},
}

@article{wang2024survey,
  author       = {Lei Wang and
                  Chen Ma and
                  Xueyang Feng and
                  Zeyu Zhang and
                  Hao Yang and
                  Jingsen Zhang and
                  Zhiyuan Chen and
                  Jiakai Tang and
                  Xu Chen and
                  Yankai Lin and
                  Wayne Xin Zhao and
                  Zhewei Wei and
                  Jirong Wen},
  title        = {A survey on large language model based autonomous agents},
  journal      = {Frontiers Comput. Sci.},
  volume       = {18},
  number       = {6},
  pages        = {186345},
  year         = {2024},
  doi          = {10.1007/S11704-024-40231-1},
}

@inproceedings{wei2023jailbroken,
  author       = {Alexander Wei and
                  Nika Haghtalab and
                  Jacob Steinhardt},
  title        = {Jailbroken: How Does {LLM} Safety Training Fail?},
  booktitle    = {NeurIPS},
  year         = {2023},
}

@article{zou2023universal,
  author       = {Andy Zou and
                  Zifan Wang and
                  J. Zico Kolter and
                  Matt Fredrikson},
  title        = {Universal and Transferable Adversarial Attacks on Aligned Language
                  Models},
  journal      = {CoRR},
  volume       = {abs/2307.15043},
  year         = {2023},
  doi          = {10.48550/ARXIV.2307.15043},
  eprinttype    = {arXiv},
  eprint       = {2307.15043},
}

@inproceedings{liu2024autodan,
  author       = {Xiaogeng Liu and
                  Nan Xu and
                  Muhao Chen and
                  Chaowei Xiao},
  title        = {AutoDAN: Generating Stealthy Jailbreak Prompts on Aligned Large Language
                  Models},
  booktitle    = {ICLR},
  publisher    = {OpenReview.net},
  year         = {2024},
}

@article{yi2024jailbreak,
  author       = {Sibo Yi and
                  Yule Liu and
                  Zhen Sun and
                  Tianshuo Cong and
                  Xinlei He and
                  Jiaxing Song and
                  Ke Xu and
                  Qi Li},
  title        = {Jailbreak Attacks and Defenses Against Large Language Models: {A}
                  Survey},
  journal      = {CoRR},
  volume       = {abs/2407.04295},
  year         = {2024},
  doi          = {10.48550/ARXIV.2407.04295},
  eprinttype    = {arXiv},
  eprint       = {2407.04295},
}

@inproceedings{zhan2025adaptive,
  author       = {Qiusi Zhan and
                  Richard Fang and
                  Henil Shalin Panchal and
                  Daniel Kang},
  title        = {Adaptive Attacks Break Defenses Against Indirect Prompt Injection
                  Attacks on {LLM} Agents},
  booktitle    = {{NAACL}},
  series       = {Findings of {ACL}},
  volume       = {{NAACL} 2025},
  pages        = {7101--7117},
  publisher    = {Association for Computational Linguistics},
  year         = {2025},
  doi          = {10.18653/V1/2025.FINDINGS-NAACL.395},
}

@article{perez2022ignore,
  author       = {F{\'{a}}bio Perez and
                  Ian Ribeiro},
  title        = {Ignore Previous Prompt: Attack Techniques For Language Models},
  journal      = {CoRR},
  volume       = {abs/2211.09527},
  year         = {2022},
  doi          = {10.48550/ARXIV.2211.09527},
  eprinttype    = {arXiv},
  eprint       = {2211.09527},
}

@misc{Kent_2025,
	author = {Sarah Kent},
	title = {{P}rompt {I}njection: {T}he {A}{I} {V}ulnerability {W}e {S}till {C}an’t {F}ix},
	howpublished = {\url{https://www.guidepointsecurity.com/blog/prompt-injection-the-ai-vulnerability-we-still-cant-fix/}},
	year={2025}, 
    month={August}
}

@inproceedings{hsieh2024found,
  author       = {Cheng{-}Yu Hsieh and
                  Yung{-}Sung Chuang and
                  Chun{-}Liang Li and
                  Zifeng Wang and
                  Long T. Le and
                  Abhishek Kumar and
                  James R. Glass and
                  Alexander Ratner and
                  Chen{-}Yu Lee and
                  Ranjay Krishna and
                  Tomas Pfister},
  title        = {Found in the middle: Calibrating Positional Attention Bias Improves
                  Long Context Utilization},
  booktitle    = {ACL},
  series       = {Findings of {ACL}},
  volume       = {{ACL} 2024},
  pages        = {14982--14995},
  publisher    = {Association for Computational Linguistics},
  year         = {2024},
  doi          = {10.18653/V1/2024.FINDINGS-ACL.890},
}

@article{liu2024lost,
  author       = {Nelson F. Liu and
                  Kevin Lin and
                  John Hewitt and
                  Ashwin Paranjape and
                  Michele Bevilacqua and
                  Fabio Petroni and
                  Percy Liang},
  title        = {Lost in the Middle: How Language Models Use Long Contexts},
  journal      = {Trans. Assoc. Comput. Linguistics},
  volume       = {12},
  pages        = {157--173},
  year         = {2024},
  doi          = {10.1162/TACL\_A\_00638},
}

@inproceedings{zhu2025melon,
  author       = {Kaijie Zhu and
                  Xianjun Yang and
                  Jindong Wang and
                  Wenbo Guo and
                  William Yang Wang},
  title        = {{MELON:} Provable Defense Against Indirect Prompt Injection Attacks
                  in {AI} Agents},
  booktitle    = {ICML},
  publisher    = {OpenReview.net},
  year         = {2025},
}

@inproceedings{yao2022react,
  author       = {Shunyu Yao and
                  Jeffrey Zhao and
                  Dian Yu and
                  Nan Du and
                  Izhak Shafran and
                  Karthik R. Narasimhan and
                  Yuan Cao},
  title        = {ReAct: Synergizing Reasoning and Acting in Language Models},
  booktitle    = {ICLR},
  publisher    = {OpenReview.net},
  year         = {2023},
}

@inproceedings{xie2024osworld,
  author       = {Tianbao Xie and
                  Danyang Zhang and
                  Jixuan Chen and
                  Xiaochuan Li and
                  Siheng Zhao and
                  Ruisheng Cao and
                  Toh Jing Hua and
                  Zhoujun Cheng and
                  Dongchan Shin and
                  Fangyu Lei and
                  Yitao Liu and
                  Yiheng Xu and
                  Shuyan Zhou and
                  Silvio Savarese and
                  Caiming Xiong and
                  Victor Zhong and
                  Tao Yu},
  title        = {OSWorld: Benchmarking Multimodal Agents for Open-Ended Tasks in Real
                  Computer Environments},
  booktitle    = {NeurIPS},
  year         = {2024},
}

@inproceedings{debenedetti2024agentdojo,
  author       = {Edoardo Debenedetti and
                  Jie Zhang and
                  Mislav Balunovic and
                  Luca Beurer{-}Kellner and
                  Marc Fischer and
                  Florian Tram{\`{e}}r},
  title        = {AgentDojo: {A} Dynamic Environment to Evaluate Prompt Injection Attacks
                  and Defenses for {LLM} Agents},
  booktitle    = {NeurIPS},
  year         = {2024},
}

@inproceedings{gur2024a,
  author       = {Izzeddin Gur and
                  Hiroki Furuta and
                  Austin V. Huang and
                  Mustafa Safdari and
                  Yutaka Matsuo and
                  Douglas Eck and
                  Aleksandra Faust},
  title        = {A Real-World WebAgent with Planning, Long Context Understanding, and
                  Program Synthesis},
  booktitle    = {ICLR},
  publisher    = {OpenReview.net},
  year         = {2024},
}

@inproceedings{kai2023not,
  author       = {Sahar Abdelnabi and
                  Kai Greshake and
                  Shailesh Mishra and
                  Christoph Endres and
                  Thorsten Holz and
                  Mario Fritz},
  title        = {Not What You've Signed Up For: Compromising Real-World LLM-Integrated
                  Applications with Indirect Prompt Injection},
  booktitle    = {AISec},
  pages        = {79--90},
  publisher    = {{ACM}},
  year         = {2023},
  doi          = {10.1145/3605764.3623985},
}

@misc{openai2025atlas,
    author = {{OpenAI}},
	title = {{I}ntroducing {C}hat{G}{P}{T} {A}tlas},
	howpublished = {\url{https://openai.com/index/introducing-chatgpt-atlas/}},
    year         = {2025},
    month        = {October},
}

@inproceedings{lefeuvre2022flexos,
  author       = {Hugo Lefeuvre and
                  Vlad{-}Andrei Badoiu and
                  Alexander Jung and
                  Stefan Lucian Teodorescu and
                  Sebastian Rauch and
                  Felipe Huici and
                  Costin Raiciu and
                  Pierre Olivier},
  title        = {FlexOS: towards flexible {OS} isolation},
  booktitle    = {ASPLOS},
  pages        = {467--482},
  publisher    = {{ACM}},
  year         = {2022},
  doi          = {10.1145/3503222.3507759},
}

@misc{packer2023memgpt,
  author       = {Charles Packer and
                  Vivian Fang and
                  Shishir G. Patil and
                  Kevin Lin and
                  Sarah Wooders and
                  Joseph E. Gonzalez},
  title        = {MemGPT: Towards LLMs as Operating Systems},
  journal      = {CoRR},
  volume       = {abs/2310.08560},
  year         = {2023},
  doi          = {10.48550/ARXIV.2310.08560},
  eprinttype    = {arXiv},
  eprint       = {2310.08560},
}

@book{betts2013cqrs,
author = {Betts, Dominic and Dominguez, Julian and Melnik, Grigori and Simonazzi, Fernando and Subramanian, Mani},
title = {Exploring CQRS and Event Sourcing: A journey into high scalability, availability, and maintainability with Windows Azure},
year = {2013},
isbn = {1621140164},
publisher = {Microsoft patterns \& practices},
edition = {1st},
}

@inproceedings{zhang2025agent,
  author       = {Hanrong Zhang and
                  Jingyuan Huang and
                  Kai Mei and
                  Yifei Yao and
                  Zhenting Wang and
                  Chenlu Zhan and
                  Hongwei Wang and
                  Yongfeng Zhang},
  title        = {Agent Security Bench {(ASB):} Formalizing and Benchmarking Attacks
                  and Defenses in LLM-based Agents},
  booktitle    = {ICLR},
  publisher    = {OpenReview.net},
  year         = {2025},
}

@article{nasr2025attacker,
  author       = {Milad Nasr and
                  Nicholas Carlini and
                  Chawin Sitawarin and
                  Sander V. Schulhoff and
                  Jamie Hayes and
                  Michael Ilie and
                  Juliette Pluto and
                  Shuang Song and
                  Harsh Chaudhari and
                  Ilia Shumailov and
                  Abhradeep Thakurta and
                  Kai Yuanqing Xiao and
                  Andreas Terzis and
                  Florian Tram{\`{e}}r},
  title        = {The Attacker Moves Second: Stronger Adaptive Attacks Bypass Defenses
                  Against Llm Jailbreaks and Prompt Injections},
  journal      = {CoRR},
  volume       = {abs/2510.09023},
  year         = {2025},
  doi          = {10.48550/ARXIV.2510.09023},
  eprinttype    = {arXiv},
  eprint       = {2510.09023},
}

@article{ferrag2025llm,
  author       = {Mohamed Amine Ferrag and
                  Norbert Tihanyi and
                  M{\'{e}}rouane Debbah},
  title        = {From {LLM} Reasoning to Autonomous {AI} Agents: {A} Comprehensive
                  Review},
  journal      = {CoRR},
  volume       = {abs/2504.19678},
  year         = {2025},
  doi          = {10.48550/ARXIV.2504.19678},
  eprinttype    = {arXiv},
  eprint       = {2504.19678},
}

@misc{protectai,
  author = {ProtectAI.com},
  title = {Fine-Tuned DeBERTa-v3-base for Prompt Injection Detection},
  year = {2024},
  publisher = {HuggingFace},
  url = {https://huggingface.co/ProtectAI/deberta-v3-base-prompt-injection-v2},
}

@misc{meta_2025_llama,
	author = {Meta},
	title = {{L}lama {P}rompt {G}uard 2 | {M}odel {C}ards and {P}rompt formats},
	howpublished = {\url{https://www.llama.com/docs/model-cards-and-prompt-formats/prompt-guard/}},
	year = {2025},
}

@inproceedings{PIGuard,
  author       = {Hao Li and
                  Xiaogeng Liu and
                  Ning Zhang and
                  Chaowei Xiao},
  title        = {PIGuard: Prompt Injection Guardrail via Mitigating Overdefense for
                  Free},
  booktitle    = {ACL},
  pages        = {30420--30437},
  publisher    = {Association for Computational Linguistics},
  year         = {2025},
}

@inproceedings{chao2025jailbreaking,
  author       = {Patrick Chao and
                  Alexander Robey and
                  Edgar Dobriban and
                  Hamed Hassani and
                  George J. Pappas and
                  Eric Wong},
  title        = {Jailbreaking Black Box Large Language Models in Twenty Queries},
  booktitle    = {SaTML},
  pages        = {23--42},
  publisher    = {{IEEE}},
  year         = {2025},
  doi          = {10.1109/SATML64287.2025.00010},
}

@misc{qwen2025qwen25technicalreport,
  author       = {An Yang and
                  Baosong Yang and
                  Beichen Zhang and
                  Binyuan Hui and
                  Bo Zheng and
                  Bowen Yu and
                  Chengyuan Li and
                  Dayiheng Liu and
                  Fei Huang and
                  Haoran Wei and
                  Huan Lin and
                  Jian Yang and
                  Jianhong Tu and
                  Jianwei Zhang and
                  Jianxin Yang and
                  Jiaxi Yang and
                  Jingren Zhou and
                  Junyang Lin and
                  Kai Dang and
                  Keming Lu and
                  Keqin Bao and
                  Kexin Yang and
                  Le Yu and
                  Mei Li and
                  Mingfeng Xue and
                  Pei Zhang and
                  Qin Zhu and
                  Rui Men and
                  Runji Lin and
                  Tianhao Li and
                  Tingyu Xia and
                  Xingzhang Ren and
                  Xuancheng Ren and
                  Yang Fan and
                  Yang Su and
                  Yichang Zhang and
                  Yu Wan and
                  Yuqiong Liu and
                  Zeyu Cui and
                  Zhenru Zhang and
                  Zihan Qiu},
  title        = {Qwen2.5 Technical Report},
  journal      = {CoRR},
  volume       = {abs/2412.15115},
  year         = {2024},
  doi          = {10.48550/ARXIV.2412.15115},
  eprinttype    = {arXiv},
  eprint       = {2412.15115},
}

@misc{openaiGPT4oMini,
    author = {{OpenAI}},
	title = {{G}{P}{T}-4o mini: advancing cost-efficient intelligence},
	howpublished = {\url{https://openai.com/index/gpt-4o-mini-advancing-cost-efficient-intelligence/}},
	year = {2024},
    month = {July},
}

@misc{openaiHelloGPT4o,
    author = {{OpenAI}},
	title = {{H}ello {G}{P}{T}-4o},
	howpublished = {\url{https://openai.com/index/hello-gpt-4o/}},
	year = {2024},
    month = {May},
}

@misc{openaiGPT51Smarter,
    author = {{OpenAI}},
	title = {{G}{P}{T}-5.1: {A} smarter, more conversational {C}hat{G}{P}{T}},
	howpublished = {\url{https://openai.com/index/gpt-5-1/}},
	year = {2025},
    month = {November}
}

@misc{anthropicClaudeSonnet,
	author = {Anthropic},
	title = {{C}laude 3.7 {S}onnet and {C}laude {C}ode},
	howpublished = {\url{https://www.anthropic.com/news/claude-3-7-sonnet}},
	year = {2025},
    month = {February}
}

@article{comanici2025gemini,
  author       = {Gemini Team},
  title        = {Gemini 2.5: Pushing the Frontier with Advanced Reasoning, Multimodality,
                  Long Context, and Next Generation Agentic Capabilities},
  journal      = {CoRR},
  volume       = {abs/2507.06261},
  year         = {2025},
  doi          = {10.48550/ARXIV.2507.06261},
  eprinttype    = {arXiv},
  eprint       = {2507.06261},
}

@misc{sandwich_defense,
  author = {Schulhoff, Sander},
  title = {The Sandwich Defense: Strengthening AI Prompt Security},
  url = {https://learnprompting.org/docs/prompt_hacking/defensive_measures/sandwich_defense},
  year = {2024},
  organization = {Learnprompting.org}
}

@article{codex,
  author       = {Mark Chen and
                  Jerry Tworek and
                  Heewoo Jun and
                  Qiming Yuan and
                  Henrique Pond{\'{e}} de Oliveira Pinto and
                  Jared Kaplan and
                  Harri Edwards and
                  Yuri Burda and
                  Nicholas Joseph and
                  Greg Brockman and
                  Alex Ray and
                  Raul Puri and
                  Gretchen Krueger and
                  Michael Petrov and
                  Heidy Khlaaf and
                  Girish Sastry and
                  Pamela Mishkin and
                  Brooke Chan and
                  Scott Gray and
                  Nick Ryder and
                  Mikhail Pavlov and
                  Alethea Power and
                  Lukasz Kaiser and
                  Mohammad Bavarian and
                  Clemens Winter and
                  Philippe Tillet and
                  Felipe Petroski Such and
                  Dave Cummings and
                  Matthias Plappert and
                  Fotios Chantzis and
                  Elizabeth Barnes and
                  Ariel Herbert{-}Voss and
                  William Hebgen Guss and
                  Alex Nichol and
                  Alex Paino and
                  Nikolas Tezak and
                  Jie Tang and
                  Igor Babuschkin and
                  Suchir Balaji and
                  Shantanu Jain and
                  William Saunders and
                  Christopher Hesse and
                  Andrew N. Carr and
                  Jan Leike and
                  Joshua Achiam and
                  Vedant Misra and
                  Evan Morikawa and
                  Alec Radford and
                  Matthew Knight and
                  Miles Brundage and
                  Mira Murati and
                  Katie Mayer and
                  Peter Welinder and
                  Bob McGrew and
                  Dario Amodei and
                  Sam McCandlish and
                  Ilya Sutskever and
                  Wojciech Zaremba},
  title        = {Evaluating Large Language Models Trained on Code},
  journal      = {CoRR},
  volume       = {abs/2107.03374},
  year         = {2021},
  url          = {https://arxiv.org/abs/2107.03374},
  eprinttype    = {arXiv},
  eprint       = {2107.03374},
  bibsource    = {dblp computer science bibliography, https://dblp.org}
}
